\documentclass[11pt]{article}
\usepackage{amsmath,amssymb}
\usepackage[dvips]{epsfig}
\usepackage[dvips]{color}
\usepackage{pb-diagram,lamsarrow,pb-lams}
\DeclareFontFamily{OT1}{rsfs}{}
\DeclareFontShape{OT1}{rsfs}{m}{n}{ <-7> rsfs5 <7-10> rsfs7 <10-> rsfs10}{}
\DeclareMathAlphabet\mathcurl{OT1}{rsfs}{m}{n}
\def\1#1{{\bf #1}}
\def\2#1{{\cal #1}}\def\9#1{{\sl #1}}\def\4#1{{\tt #1}}\def\5#1{{\sf #1}}
\def\6#1{{\mathfrak #1}}\def\7#1{{\mathbb #1}}\def\8#1{{\rm #1}}
\def\9#1{{\mathcurl #1}}

\def\3{{\ss}}

\def\skb{\vskip 0.5cm}

\def\beq{\begin{eqnarray}}
\def\eeq{\end{eqnarray}}
\def\vs{\vspace{0.2cm} \\}

\def\Poin{{\8P^\uparrow_+}}

\newtheorem{The}{Theorem}[section]
\newtheorem{Def}[The]{Definiton}
\newtheorem{Lem}[The]{Lemma}
\newtheorem{Pro}[The]{Proposition}
\newtheorem{Cor}[The]{Corollary}
\newtheorem{Slem}[The]{Sublemma}
\def\bdef{\begin{Def}\1: \em}
\def\eef{\end{Def}}

\def\blem{\begin{Lem}\1: }
\def\elem{\end{Lem}}
\def\bslem{\begin{Slem}\1: }
\def\eslem{\end{Slem}}
\def\bthe{\begin{The}\1: }
\def\ethe{\end{The}}
\def\bpro{\begin{Pro}\1: }
\def\epro{\end{Pro}}
\def\bcor{\begin{Cor}\1: }
\def\ecor{\end{Cor}}

\def\al{\alpha}
\def\be{\beta}
\def\gam{\gamma}\def\Gam{\Gamma}
\def\lam{\lambda}\def\Lam{\Lambda}
\def\eps{\epsilon} 
\def\te{\theta}

\def\sgm{\sigma}
\def\om{\omega}\def\Om{\Omega}

\def\bpr{\paragraph*{\it Proof.}}

\def\epr{$\square$\skb}

\def\pa{\partial}
\def\<{\langle}
\def\>{\rangle}
\def\Ad{{\rm Ad}}

\def\bs{\backslash}

\def\bdes{\begin{enumerate}}
\def\edes{\end{enumerate}}
\newcommand\itno[1]{\item[{\it ({#1})}]}

\def\bmat{\left( \begin{array}{ccc} }
\def\emat{\end{array} \right)}

\def\beqa{\begin{eqnarray*}}
\def\eeqa{\end{eqnarray*}}
\def\bdia{\begin{diagram}}
\def\edia{\end{diagram}}

\def\ul{\underline}

\def\rMapsto{ \ \longmapsto \ }
 
\title{\bf Short-distance analysis 
for algebraic euclidean field theory}
\author{{\it Dirk Schlingemann} \\
The Erwin Schr\"odinger International Institute \\ 
for Mathematical Physics (ESI)\\
Vienna}
\begin{document}
\maketitle
\abstract{Recently D. Buchholz and R. Verch have proposed a method for 
implementing in algebraic quantum field theory ideas from 
renormalization group analysis of short-distance (high energy) 
behavior by passing to certain scaling limit theories. 
Buchholz and Verch distinguish between different types of theories where the 
limit is unique, degenerate, or classical, and the method allows in 
principle to extract the `ultraparticle' content of a given model, 
i.e. to identify particles (like quarks and gluons) that are not 
visible at finite distances due to `confinement'.  It is therefore of 
great importance for the physical interpretation of the theory.  The 
method has been illustrated in a simple model in with some 
rather surprising results.

This paper will focus on the question how the short distance 
behavior of models defined by euclidean means is reflected in the 
corresponding behavior of their Minkowski counterparts.  More 
specifically, we shall prove that if a euclidean theory has some 
short distance limit, then it is possible to pass from this limit 
theory to a theory on Minkowski space, which is a short distance limit 
of the Minkowski space theory corresponding to the original euclidean 
theory.}
\newpage
\section{Introduction}
In the past three decades several approaches have been developed that 
incorporate the physical principles of quantum field theory (QFTh) in a 
mathematically rigorous fashion.  Among these rigorous approaches the 
framework of algebraic quantum field theory is probably the 
most highly developed. Its basic objects are algebras of observables, 
indexed by domains of space-time where the observables can be 
measured.
Hence the interpretation of the basic structure is rather 
clear-cut.
Many deep, model independent results have been obtained within this 
framework,
which has 
also proved well adapted to the construction and analysis of models in 
two dimensional conformal field theory. 

On the other hand, models of quantum fields based on concrete 
lagrangians are usually constructed by means of euclidean functional 
integrals \cite{GlJa4}.  
From the constructive point of view these methods have 
many advantages, for the basic algebraic structure is commutative and 
powerful tools from classical statistical mechanics can be applied.  
The passage from such objects in euclidean space to algebraic quantum 
field theory on Minkowski space time is mathematically a highly 
nontrivial operation, however, and it is by no means clear in general 
how properties of the former are reflected in latter.  Hence the 
physical interpretation is much less obvious than in algebraic quantum 
field theory.  When the euclidean theory has only been proved to exist 
in a bounded euclidean domain, as is the case for the four dimensional 
Yang-Mills model constructed by Magnen, Rivasseau and Sen\'eor 
\cite{MagRivSen93}, the passage itself is an important unsolved problem.

Recently D. Buchholz and R. Verch have proposed a method for 
implementing in algebraic quantum field theory ideas from 
renormalization group analysis of short-distance (high energy) 
behavior by passing to certain scaling limit theories 
\cite{Bu94,BuVer95,Bu96b,Bu96a,BuVer97,Bu97}.  
Buchholz and Verch distinguish between different types of theories where the 
limit is unique, degenerate, or classical, and the method allows in 
principle to extract the `ultraparticle' content of a given model, 
i.e. to identify particles (like quarks and gluons) that are not 
visible at finite distances due to `confinement'.  It is therefore of 
great importance for the physical interpretation of the theory.  The 
method has been illustrated in a simple model in \cite{BuVer97} with some 
rather surprising results.

This paper will focus on the question how the short distance 
behavior of models defined by euclidean means is reflected in the 
corresponding behavior of their Minkowski counterparts.  More 
specifically, we shall prove that if a euclidean theory has some 
short distance limit, then it is possible to pass from this limit 
theory to a theory on Minowski space, which is a short distance limit 
of the Minkowski space theory corresponding to the original euclidean 
theory.

\paragraph{\em The present status of algebraic euclidean field theory.}
The techniques of euclidean field theory (EFTh)
\cite{GlJa4} have proved to be very 
powerful for the construction of interacting quantum field theory 
models and often superior to the method of canonical quantization in 
Minkowski space.  For instance, existence of the $\phi^4_3$ model as a 
Wightman quantum field theory has been established by using euclidean 
methods \cite{FeldOst,SeilSim76,MagSen76} combined with the 
Osterwalder-Schrader reconstruction theorem \cite{OstSchra1}.  Within 
a hamiltonian framework essentially only the proof of the positivity 
of the energy has been carried out.  Also in cases where a direct 
Minkowski space construction is possible, as in the $P(\phi)_2$ and 
Yukawa$_2$ models, euclidean techniques may simplify things 
considerably, e.g. in in the proof of of Poincar\'e covariance or 
discussions of phase transitions and symmetry breaking.  In these 
constructions the key objects are usually the euclidean Greens 
functions, or Schwinger distributions, $\6S_n$ that are represented as 
moments of a measure $d\mu$ on a space of distributions $S'(\7R^d)$ 
\beqa 
\6S_n(x_1,\cdots,x_n)=\int\8d\mu(\phi) \ \phi(x_1)\cdots\phi(x_n). 
\eeqa 
Methods from statistical mechanics like renormalization group 
analysis \cite{GawKup} and cluster expansions \cite{Br} can be applied 
in order to perform the continuum and the infinite volume limits of 
lattice regularized models.  But the construction of Schwinger 
distributions is not enough, the problem of linking them to 
physics in Minkowski space has to be addressed.

The Osterwalder-Schrader reconstruction theorems \cite{OstSchra1} connect the 
Schwinger distributions of an euclidean field theory with the Wightman 
distributions of a quantum field theory on Minkowski space.  Powerful 
as these theorems are, there are several reasons why they can not be 
considered as the final answer to the problem of linking euclidean and 
Minkowski space theories.  For one thing, the conditions of the 
equivalence theorem (Theorem $E\leftrightarrow R$  in \cite{OstSchra1}) 
are extremely hard to verify, 
while the convenient sufficient conditions for the passage from 
Schwinger distributions to Wightman distributions 
(Theorem $E'$(or $E''$)$\rightarrow R'$  in \cite{OstSchra1}) 
are most probably too restrictive in general.  Secondly, and this is a 
more important point than the first for the present discussion, the 
results in \cite{OstSchra1} do not allow one to conclude that a local net of 
observable algebras in the sense of algebraic quantum field theory can 
be obtained from the Schwinger distributions.  In fact, the question 
when a Wightman quantum field gives rise to such a net is quite a 
delicate one, see \cite{BorYng} for a review.  Useful sufficient 
conditions are known, however \cite{GlJa4,DrieFroh77}.  
In the models with point fields 
constructed so far these conditions are fulfilled, but it appears very 
unlikely that they will be so in general.

The third point is that Schwinger distributions, which are euclidean 
expectation values of point fields, may not be adequate in gauge 
theories and one should rather consider expectation values of extended 
objects localized around loops or even strings extending to infinity. 
An Osterwalder-Schrader-type reconstruction scheme
which can be applied to correlation functions of loops and strings 
was established by E. Seiler \cite{Seil82} and 
J. Fr\"ohlich, K. Osterwalder and E. Seiler \cite{FrohOstSeil}. Local 
commutativity of the reconstructed observables remained an open 
problem in this work, however. 

A C*-algebraic version of the  reconstruction 
theorems in \cite{OstSchra1,DrieFroh77,GlJa4}, 
which to a certain extent generalizes the previous 
considerations in \cite{FrohOstSeil,Seil82} and solves the locality 
problem, has been worked out in \cite{Schl97}.  The starting point of 
this analysis is a net $\6B$ of C*-algebras, indexed by regions in 
euclidean space, and acted upon by an action $\gam$ of the euclidean 
group by automorphisms of $\6E$.  The third ingredient is a continuous 
euclidean invariant and reflexion positive \cite{GlJa4} 
functional $\eta$ on $\6B$. It is shown in 
\cite{Schl97} that for a given euclidean field 
$(\6B,\be,\eta)$ a Haag-Kastler net $\6A$ on Minkowski space, 
a covariant action of the Poincar\'e group by 
automorphisms $\al$ on $\6A$ as well as a vacuum state 
$\om$ on $\6A$ can directly be reconstructed from 
$(\6B,\be,\eta)$.  The main advantage of the C*-algebraic framework, 
is that one only deals with bounded operators from the outset.  This 
is also important for the proof of locality for the 
constructed Haag-Kastler net. Finally, we mention that
there are indications that a C*-algebraic point of view 
also enlarges the variety of constructible euclidean field theory 
models \cite{Schl98}. 

\paragraph{\em Scaling algebras and renormalization group.}
A general approach for the analysis of the high energy properties
of a given quantum field theory model has been developed by 
D. Buchholz and R. Verch \cite{Bu94,BuVer95,Bu96b,Bu96a,BuVer97,Bu97}. 
The starting point of their analysis is a 
quantum field theory formulated within the 
C*-algberaic approach as it has been introduced by
R. Haag and D. Kastler \cite{H,HK}.  
We briefly recall here the mathematical description of 
this framework.  

A $\Poin$-covariant Haag-Kastler net is an 
inclusion preserving prescription 
which assigns to each double cone 
$\9O=V_++x\cap V_-+y$ a unital C*-subalgebra $\6A(\9O)\subset\6A$
of a C*-algebra $\6A$.  
\footnote{Here $V_\pm=\{x|x^2>0,\pm x^0>0\}$ is the forward (backward)
light cone in Minkowski space.} The self-adjoint elements
within a local  algebra $\6A(\9O)$ correspond to 
observables which can be measured within the spacetime 
region $\9O$. 
The Poincar\' e group $\Poin$ 
acts covariantly on the net $\9O\mapsto\6A(\9O)$, i.e. there is a 
group homomorphism $\al$ from the Poincar\'e group into the automorphism
group of $\6A$, such that $\al_g\6A(\9O)=\6A(g\9O)$
for each Poincar\'e transformation 
$g$. The concept of locality
is encoded by the property 
that, if $\9O,\9O_1$ are two
spacelike separated regions, then 
the operators in $\6A(\9O)$ commute with those in $\6A(\9O_1)$, i.e.
two measurements which are performed in spacelike separated regions
are commensurable.  
We write for the corresponding $\Poin$-covariant Haag-Kastler net
$(\6A,\al)$. 

One selection criterion for physical states, which is related to 
a stability requirement, is the so called 
{\em spectrum condition}. A state $\om$, ehich is subject to this condition,
is called a {\em positive energy state}, 
characterized by the property that there 
exists a unitary strongly continuous 
representation $U$ of the Poincar\'e group 
on the GNS Hilbert space of $\om$, implementing the automorphisms 
$\al_g$ in the GNS representation of $\om$,
such that the spectrum of the generator of the translation group 
$x\mapsto U(x)$ 
is contained in the closed forward light cone $\bar V_+$. 
A particular class of  positive energy states are the {\em vacuum 
states}, which also have the property to be {\em Poincar\'e invariant},
$\om\circ\al_g=\om$. This reflects the fact that 
within a vacuum there is no matter configuration which 
can distinguish a certain region in spacetime. 

A triple $(\6A,\al,\om)$, where $(\6A,\al)$ is a Haag-Kastler net
and $\om$ is a positive energy state is called a {\em quantum field}. 

The concept of scaling algebra allows to express some basic ideas of 
renormalization group analysis within the algebraic framework. 
For a positive number $\lam>0$ one builds a new Haag-Kastler net 
$\9O\mapsto\6A_\lam(\9O)$ by defining the scaled algebra of a 
domain $\9O$ by
$\6A_\lam(\9O):=\6A(\lam\9O)$ and putting
$\al_{(\lam,g)}:=\al_{\lam \circ g \circ \lam^{-1}}$ for  
a Poincar\'e transformation $g$.
Thus one keeps Minkowski space fixed and one interprets the properties 
of the given theory at small scales
(high energy behavior) in terms of the modified theories
$\9O\mapsto \6A_\lam(\9O)$.
 
The {\em scaling algebra} $\ul{\6A}$  is the C*-algebra 
which is generated by a certain class of bounded functions  
$\1a:\7R_+\to\6A$ (see \cite{BuVer97} and related work).  
The functions in $\ul{\6A}$ are regarded as 
{\em orbits} under the action of the
renormalization group transformations which identify 
operators at scale $1$ with operators at scale $\lam$. 
This requires a particular scaling behavior in configuration space, 
namely an operator $\1a\in\ul{\6A}$ is localized in 
$\9O$ if for each $\lam\in \7R_+$ the operator 
$\1a(\lam)$ is localized in the scaled region $\lam\9O$.
On the other hand, a condition for the 
scaling behavior in momentum space
is needed in order to fix Planck's constant $\hbar$.
Formulated in terms of the scaling algebra, the 
correct scaling in momentum space can be archieved by 
requiering that the continuety property 
\beqa
\lim_{g\to 1}\sup_{\lam\in\7R_+}
\|\1a(\lam)-\al_{\lam\circ g\circ\lam^{-1}}\1a(\lam)\|&=&0
\eeqa
is fulfilled for each $\1a\in\ul{\6A}$. In other words, the 
group homomorphism $\ul{\al}$ from the Poincar\'e group 
into the automorphism group of $\ul{\6A}$, which is 
given by 
\beqa
(\ul{\al}_g\1a)(\lam)&:=&\al_{\lam\circ g\circ\lam^{-1}}\1a(\lam)
\eeqa
is strongly continuous. 
  
Each physical state $\om$ (in particular a vacuum state) 
of the underlying theory 
$\9O\mapsto \6A(\9O)$ can be lifted to a physical state $\ul{\om}$ on the 
scaling algebra by the prescription
\beqa 
\<\ul{\om},\1a\>&:=&\<\om,\1a(1)\> \ \ .
\eeqa
Hence the state $\ul{\om}$ evaluates the  
renormalization group orbit $\1a$ at scale $\lam=1$.
 
The group of scaling transformations $\sgm_\lam$ acts by 
automorphism on the scaling algebra in a natural fashion  
\beqa
(\sgm_{\lam}\1a)(\lam_1)&:=&\1a(\lam_1\lam)  \ \ ,
\eeqa
leading to a net of states 
\beqa
\{\ul{\om}_\lam=\ul{\om}\circ\sgm_\lam|0<\lam\} 
\eeqa
which has, according to the weak-compactness of the set of states 
of a C*-algebra,  weak limit points $\ul{\om}_\zeta$ for $\lam\to 0$ 
(the symbol $\zeta$ labels such a limit point).
It has been proven in \cite{BuVer95,BuVer97} 
that all weak limit points $\ul{\om}_\zeta$ 
are vacuum states and for each of them 
one obtains a quantum field  
$(\ul{\6A}_\zeta,\ul{\al}_\zeta,\ul{\om}_\zeta)$, where 
the algebra $\ul{\6A}_\zeta$ is defined by
\beqa
\ul{\6A}_\zeta&:=&\ul{\6A}/\ul{\pi}_\zeta^{-1}(0) 
\eeqa
where $\ul{\pi}_\zeta$ is the GNS representation 
of $\ul{\om}_\zeta$. The group homomorphism 
$\ul{\al}_\zeta$ is the lifting of $\ul{\al}$ to $\ul{\6A}_\zeta$,
which exists, since the ideal $\ul{\pi}_\zeta^{-1}(0)$ is Poincar\'e
invariant.
These quantum fields are called 
{\em scaling limits} of the quantum field $(\6A,\al,\om)$
and they describe the high energy 
behavior of the underlying theory.  
D. Buchholz and R. Verch distinguish three cases in order 
to classify the scaling limits: 
\bdes 
\itno 1
All scaling limit theories are equivalent, i.e.
the scaling limit is unique. 
\itno 2
All scaling limit theories are just multiples of the identity, i.e.
one obtains a classical scaling limit.
\itno 3
Neither case {\it (1)} nor case {\it (2)} are valid, i.e. the 
scaling limit is degenerate.
\edes

\paragraph{\em Taking scaling limits and passing form EFTh to QFTh.}
For or a given euclidean field $(\6B,\be,\eta)$ 
the short-distance behavior
can be obtained by first 
passing by means of the construction procedure \cite{Schl97}, which
we are going to explain in 
Section \ref{sc1} in more detail, to the corresponding
quantum field theory model $(\6A,\al,\om)$ and 
in a second step applying the
analysis of D. Buchholz and R. Verch \cite{BuVer95,BuVer97} in order to get  
the scaling limit theories 
$(\ul{\6A}_\zeta,\ul{\al}_\zeta,\ul{\om}_\zeta)$.
We illustrate this procedure digramatically in the following way:
\beqa
\bdia
\node{(\6B,\be,\eta)}\arrow{e,t}{\8{recon}}
\node{(\6A,\al,\om)}
\arrow{e,t}{\8{sclim}}
\node{(\ul{\6A}_\zeta,\ul{\al}_\zeta,\ul{\om}_\zeta)} 
\edia
\eeqa

The two step procedure, described above, is rather cumbersome and it is 
advantageous to be able to study the scaling limit theories directly on 
the euclidean level. More precisely, one wishes to build from a 
euclidean field $(\6B,\be,\eta)$ the {\em euclidean scaling limit 
theory} $(\ul{\6B}_{\zeta},\ul{\be}_\zeta,\ul{\eta}_\zeta)$ first and then, 
in a second step, one constructs 
the corresponding Minkowski quantum field theory model which we denote by  
$(\6A_\zeta,\al_\zeta,\om_\zeta)$, without underlining the 
the symbols.  
\beqa
\bdia
\node{(\6B,\be,\eta)}\arrow{e,t}{\8{sclim}}
\node{(\ul{\6B}_{\zeta},\ul{\be}_\zeta,\ul{\eta}_\zeta)}\arrow{e,t}{\8{recon}}
\node{(\6A_\zeta,\al_\zeta,\om_\zeta)}
\edia
\eeqa
 
In Section \ref{sc2} we describe how to build the euclidean scaling limit 
theory $(\ul{\6B}_{\zeta},\ul{\be}_\zeta,\ul{\eta}_\zeta)$
form a given euclidean field and we present there the main result of 
this paper, namely:

\paragraph{Theorem:}
{\em The quantum field theories, which can be reconstructed from 
euclidean scaling limit theories, are equivalent to scaling limit 
theories of the quantum field theory which can be reconstructed from
the underlying euclidean field theory.} 
\skb

Formally expressed in terms of diagrams, this means that the 
digram, given below, commutes in the sense of equivalence classes
of quantum fields:
\beqa
\bdia
\node{(\6B,\be,\eta)}
\arrow{e,t}{\8{sclim}}\arrow{s,l}{\8{recon}}
\node{(\ul{\6B}_{\zeta},\ul{\be}_\zeta,\ul{\eta}_\zeta)}
\arrow{s,r}{\8{recon}} \\
\node{(\6A,\al,\om)}\arrow{e,t}{\8{sclim}}
\node{(\ul{\6A}_\zeta,\ul{\al}_\zeta,\ul{\om}_\zeta)
\cong (\6A_\zeta,\al_\zeta,\om_\zeta)}
\edia
\eeqa
At this point, we briefly explain here, what equivalence
of quantum fields means within our framework: Two quantum fields 
$(\6A,\al,\om)$ and $(\hat\6A,\hat\al,\hat\om)$ are called equivalent 
if there exists an algebra isomorphism $\iota:\6A\to\hat\6A$
such that $\iota$ intertwines the group homomorphisms $\al$ and
$\hat\al$, i.e. $\iota\circ\al_g=\hat\al_g\circ\iota$ holds true for 
each Poincar\'e transformation $g$, the states 
$\om$ and $\hat\om$ are related by $\hat\om\circ\iota=\om$, 
and the isomorphism $\iota$ respects the net structure, i.e.
for each bounded and convex region $\9U\subset\7R^d$ 
the identity $\iota(\6A(\9U))=\hat\6A(\9U)$ is valid.

\section{From euclidean field theory to quantum field theory}
\label{sc1}
Within this section we briefly discuss the ideas and strategies 
which have been developed in \cite{Schl97}.
The starting point in the framework of algebraic
euclidean field theory is an isotonous net 
\beqa
\9U\rMapsto \6B(\9U)\subset \6B
\eeqa
of C*-algebras, indexed by the set 
$\9K^d$ of bounded convex regions $\9U$ in $\7R^d$, on which 
the euclidean group $\8E(d)$ acts covariantly by 
automorphisms, i.e. there exists a group homomorphism 
$\be$ from the euclidean group $\8E(d)$ into the automorphism 
group of $\6B$ such that  
\beqa
\be_g\6B(\9U)&=&\6B(g\9U)
\eeqa
for each $\9U\in\9K^d$ and for each $g\in\8E(d)$. 
In order to implement the concept of locality within the 
euclidean framework, we assume 
that two operators commute if they are localized in 
disjoint regions, i.e. if $\9U_1\cap\9U_2=\emptyset$, then
\beqa
[\6B(\9U_1),\6B(\9U_2)]&=&\{0\} \ \ .
\eeqa 

A further ingredient for building quantum field theory models from euclidean 
data are {\em reflexion positive 
euclidean invariant regular states}. These states fulfill the following 
conditions:

\paragraph{\em Euclidean invariance:}
For each euclidean transformation $h\in\8E(d)$:
$\eta\circ\be_h=\eta$.

\paragraph{\em Reflexion positivity:}
Let $e\in S^{d-1}$ be a euclidean time direction, then 
we denote by $\6B(e)$ the C*-algebra generated by operators
which are localized in the half space $\7R_+e+e^\perp$, where 
$e^\perp$ is the hyperplane, orthogonal to $e$.
A state $\eta$ is reflexion positive if the sesquilinear form
\beq\label{equrp}
\6B(e)\otimes\6B(e)\ni b_1\otimes b_2\mapsto \<\eta,j_e(b_1)b_2\>
\eeq
is positive semidefinite. Here $j_e$ is the antilinear 
involution 
\beqa
j_e(b)&=&\be_{\te_e}(b^*)
\eeqa
with $\te_e:x\mapsto -2ex+x$.  

\paragraph{\em Regularity:}
For each $b_1,b_2,b_3\in\6B$ the map 
\beqa
h\mapsto\<\eta,b_1\be_h(b_2)b_3\>
\eeqa
is continuous.

\paragraph{\em Remarks.}
\bdes
\itno 1
Without loss of generality, we may assume that the 
GNS representation $\pi_\eta$ of $\eta$ is faithful. 
Otherwise, we simply replace the algebra $\6B$ by the 
quotient $\6B/\pi_\eta^{-1}(0)$. Note that 
the automorphism $\be_h$ can be lifted to an automorphism 
on $\6B/\pi_\eta^{-1}(0)$ since $\eta$ euclidean invariant.
\itno 2
We claim here, that 
regularity is automatically fulfilled for $\eta$ 
if the group homomorphism $\be$ is strongly continuous, i.e.
$\lim_{g\to 1}\|\be_g(b)-b\|=0$ for each $b\in\6B$.
\edes 
\skb

We showed in \cite{Schl97} how to construct from 
a given euclidean field a quantum field theory in a 
particular vacuum representation. 
In order to point out the relation between the euclidean 
field $(\6B,\be,\eta)$ and the minkowskian world, 
we briefly describe the construction of a 
Hilbert space $\2H$ on which the physical 
observables are represented, the construction of a 
unitary strongly continuous representation of the 
Poincar\'e group on $\2H$, as well as the Haag-Kastler net of 
local algebras. 

\paragraph{\em Step 1:}
By dividing the null-space of the positive semidefinite sesquilinear 
form, introduced by Equation \ref{equrp}, and by taking the closure, 
we obtain a Hilbert space $\2H$. The corresponding 
canonical projection onto the quotient is denoted by 
\beqa
\Psi:\6B(e)\mapsto \2H 
\eeqa
and we write $\Om:=\Psi[\11]$.
A unitary strongly continuous representation of the Poincar\' e group 
$U$ on $\2H$ can be constructed, which works essentially analogous 
to the procedure which has been presented in \cite{FrohOstSeil}
(compare also \cite{Seil82,Schl97}). 
The vector $\Om$ is invariant under the action of $U$.
Moreover, the spectrum of the the generator of the 
translations $x\mapsto U(x)$ is contained in the 
closed forward light cone $\bar V_+$.
 
\paragraph{\em Step 2:}
The construction of a Poincar\'e covariant Haag-Kastler net of 
bounded operators on $\2H$ can be performed analogously as 
it has been carried out in \cite{Schl97}. 
We identify bounded operators on $\2H$ by making use of 
the following proposition:

\bpro\label{pro01}
For each $s\in\7R_+$ and for each $b\in\6B(e,s)$, there exists 
a bounded operator $\pi_s(b)\in\6B(\2H)$ with 
\beqa
\|\pi_s(b)\|&\leq&\|b\|
\eeqa
which is uniquely determined by the relation
\beqa
\pi_s(b)\Psi[b_1]&=&\Psi[b\be_{se}(b_1)]
\eeqa
for each $b_1\in\6B(e)$.
\epro
\bpr
The result follows by an application of the proof of 
\cite[Theorem 10.5.5]{GlJa4}.
\epr

If we assume that $\be $ is strongly continuous, we expect, however, 
that all operators in $\ul{\6B}$, 
which are localized in the time-slice $e^\perp$,
are multiples of the identity.
But time-slice operators may be found in an 
appropriate {\em extension} of the euclidean 
field $(\6B,\be,\eta)$. 

\paragraph{\em Extension of euclidean nets.}
We call an euclidean field $(\hat{\6B},\hat\be,\hat\eta)$
an {\em extension} of $(\6B,\be,\eta)$ if 
$\hat{\6B}(\9U)\supset\6B(\9U)$ holds true for each bounded and convex set 
$\9U\subset\7R^d$, and $\hat\be_h|_{\6B}=\be_h$, $\hat\eta|_{\6B}=\eta$,
for each euclidean transformation $h$.  
\skb

Indeed, there is a natural extension $(\hat{\6B},\hat\be,\hat\eta)$ 
of the euclidean field $(\6B,\be,\eta)$:
Let $(\2K,\tau,E)$ be the GNS triple of $\eta$.
We introduce a topology on $\6B$ by semi norms 
\beqa
\|b\|_\psi&:=&\|\tau(b)\psi\|
\eeqa
with $\psi\in\2K$.
We denote by $\hat\6B$ the closure of $\6B$ within this topology and 
$\hat\6B$ is a W*-algebra, isomorphic 
to the von Neumann algebra $\tau(\6B)''$.\footnote{
For a algebra $\6M\subset\6B(\2H)$, we write 
$\6M'$ for the commutatnt of $\6M$, i.e. the algebra of 
operators in $\6B(\2H)$ commuting with those in $\6M$.}
Obviously, the homomorphism $\be$ can be 
extended to a homomorphism $\hat\be$ form $\8E(d)$ into the 
automorphism group of $\hat\6B$ and $\eta$ can be 
extended to a reflexion positive euclidean invariant 
regular state $\hat\eta$ on $\hat\6B$. 
 
For a subset $\9V$ of the hyperplane $e^\perp$ we introduce the
algebra $\hat\6B(\9V)$ of {\em time zero operators} 
which is given by the intersection
\beqa
\hat\6B(\9V)&:=&\bigcap_{s\in\7R_+} \hat\6B([0,s)e\times\9V) \ \ .
\eeqa



A Hilbert space $\hat\2H$ can be constructed from 
the extended euclidean field $(\hat\6B,\hat\be,\hat\eta)$ by Step 1
and it is isomorphic to $\2H$. Hence we may identify  
both spaces in the subsequent, i.e. $\hat\2H=\2H$.
Analogously to the analysis, carried out in \cite{Schl97},
we get: 
 
\bpro\label{pro02}
There exists a *-representation $\pi$ of the time-zero algebra 
$\hat\6B(e^\perp)$ on $\2H$.
Which is uniquely determined by the relation
\beqa
\pi(b)\Psi[b_1]&=&\Psi[bb_1]
\eeqa
for each $b_1\in\hat\6B(e)$.
\epro

For a double cone $\9O$, we define $\6A(\9O)$ to be the 
C*-algebra of operators on $\2H$ which is generated by 
all operators 
\beq\label{opm1}
\Pi[f,b] &:=& \int \8dg \ f(g) \ U(g)\pi(b)U(g)^*
\eeq
with $b\in\hat\6B(\9V)$, $f\in\9C^\infty_0(\Poin)$, such that  
$g\9V\subset\9O$ for each 
Poincar\'e transformation $g$ in the support of $f$.
The prescription 
\beqa
\9O\mapsto\6A(\9O)
\eeqa 
is an isotonous net of C*-algebras and by putting 
$\al_g:=\Ad(U(g))$, for each $g\in\Poin$, and 
$\om:=\<\Om,(\cdot)\Om\>$ we obtain quantum field according to 
$\cite{Schl97}$:

\bthe\label{reconst}
Let $(\6B,\be,\eta)$ be a euclidean field.
Then the triple $(\6A,\al,\om)$, constructed above, 
is a $\Poin$-covariant quantum field.
\ethe


\section{On the Short-distance analysis of field theories}
\label{sc2}
We briefly review the concept of scaling algebras which has been 
invented by D. Buchholz and R. Verch 
\cite{Bu97,BuVer97,Bu96a,Bu96b,BuVer95,Bu94}.

\paragraph{\em Taking scaling limits.}
In order to label the scaling limits, in a 
elegant manner, we introduce here the notion {\em 
limit functional}.

Let $\9F_\8b(\7R_+)$ be C*-algebra of all bounded 
functions on $\7R_+$ and the closed two-sided ideal $\9F_0(\7R_+)$
in $\9F_\8b(\7R_+)$
which is generated by functions $f\in\9F_\8b(\7R_+)$ 
with $\lim_{\lam\to 0}f(\lam)=0$. Then we build the 
quotient C*-algebra 
\beqa
\8{C}(\7R_+)&=&\9F_\8b(\7R_+)/\9F_0(\7R_+) \ \ ,
\eeqa
the {\em corona algebra}. Writing $\8{Sp}[\6C]$ for the 
spectrum of an abelian C*-algebra $\6C$, 
the corona algebra $\8{C}(\7R_+)$ can by 
interpreted as the algebra of functions which are supported on
$\8{Sp}[\8C(\7R_+)]=\8{Sp}[\9F_\8b(\7R_+)]\bs\8{Sp}[\9F_0(\7R_+)]$.
We claim here that $\8{Sp}[\9F_0(\7R_+)]$ is not homoeomorphic to 
$\7R_+$ since $\9F_0(\7R_+)$ contains also functions which are 
discontinuous on $\7R_+$. The set of states 
$\6S[\8{C}(\7R_+)]$ on $\8{C}(\7R_+)$ are called 
{\em limit functionals} and can be identified with the set 
of states 
$\6S[\9F_\8b(\7R_+)]$ on $\9F_\8b(\7R_+)$ which annihilate the 
ideal $\9F_0(\7R_+)$. The reason why 
the states on $\8{C}(\7R_+)$ are called limit functionals 
becomes clear by looking at a function $f\in\9F_\8b(\7R_+)$ 
for which $f_0=\lim_{\lam\to 0}f(\lam)$ exists. Namely,
for each functional $\zeta\in\6S[\8{C}(\7R_+)]$
the expectation value $\<\zeta,f\>=f_0$ 
coincides with the limit of $f$ for $\lam\to 0$ since $\zeta$ vanishes on 
$f-f_0\11\in\9F_0(\7R_+)$. A limit functional, ore 
more general, a state  $\xi$ on $\9F_\8b(\7R_+)$
can be regarded as a measure on the compact Hausdorff space 
$\8{Sp}[\9F_\8b(\7R_+)]$ and we write sometimes 
\beqa
\<\xi,f\>&=&\int\8d\xi(\lam) \ f(\lam)
\eeqa
in a suggestive manner.

\paragraph{\em Scaling limits for the euclidean fields.}
The {\em scaling algebras} are given as follows: Let 
$\9F_\8b(\7R_+,\6B)$ be the C*-algebra of bounded 
$\6B$-valued functions on $\7R_+$  then the prescription
which is given according to 
\beqa
(\ul{\be}_{h}\1b)(\lam)&:=&\be_{\lam\circ h\circ\lam^{-1}}\1b(\lam) 
\eeqa
for each euclidean transformation $h$,
yields an action $\ul{\be}$ of the euclidean group
by automorphisms on $\9F_\8b(\7R_+,\6B)$.
In order to select the admissible {\em orbits} of 
renormalization group transformations in $\9F_\8b(\7R_+,\6B)$,
we consider the C*-subalgebra $\ul{\6B}$ in 
$\9F_\8b(\7R_+,\6B)$ on which $\ul{\be}$ is strongly continuous.
In particular, for each bounded convex set $\9U$ we denote by 
$\ul{\6B}(\9U)$ the C*-subalgebra in $\ul{\6B}$ which is 
generated by elements $\1b\in\ul{\6B}$ with 
$\1b(\lam)\in\6B(\lam\9U)$. 
This definition implies that $\ul{\be}$ is covariant, i.e.
$\ul{\be}_h$ maps $\ul{\6B}(\9U)$ onto $\ul{\6B}(h\9U)$.
In other words, the pair $(\ul{\6B},\ul{\be})$ is a
euclidean net of C*-algebras.

We are now prepared to build 
for a given euclidean field $(\6B,\be,\eta)$ the corresponding 
scaling limits be means of limit functionals: 
For a given limit functional $\zeta\in\6S[\8C(\7R_+)]$ 
the {\em scaling limit state} $\ul{\eta}_\zeta$ on 
the scaling algebra $\ul{\6B}$ is given 
according to the prescription 
\beqa
\<\ul{\eta}_\zeta,\1b\>&:=&\int \8d\zeta(\lam) \ \<\eta,\1b(\lam)\>
\eeqa
for $\1b\in\ul{\6B}$.
Of course, we can build the state $\ul{\eta}_\xi$ for 
any state $\xi$ on $\9F_\8b(\7R_+)$:

\bpro\label{pro1}
Let $(\6B,\be,\eta)$ be a euclidean field. Then 
for each state $\xi$ on $\9F_\8b(\7R_+)$ the 
triple $({\ul{\6B}},\ul{\be},\ul{\eta}_\xi)$ is a euclidean field, i.e. 
$\ul{\eta}_\xi$ is a reflexion positive and euclidean invariant 
regular state on the scaling algebra.
\epro
\bpr
Since $\eta$ is a euclidean invariant state,
we conclude for each $h\in\8E(d)$ and for each 
$\lam\in\7R_+$
\beqa
\<\eta,\be_{\lam\circ h\circ\lam^{-1}}\1b(\lam)\>
&=&\<\eta,\1b(\lam)\> 
\eeqa
for each element $\1b\in\ul{\6B}$ of the euclidean scaling algebra.
Applying the functional $\xi$ to both sides yields 
\beqa
\<\ul{\eta}_\xi, \ul{\be}_g\1b\>&=&
\int\8d\xi(\lam) \ \<\eta,\be_{\lam\circ g\circ\lam^{-1}}\1b(\lam)\>
\vs\vs
&=&\int\8d\xi(\lam) \ \<\eta,\1b(\lam)\>
\vs\vs
&=&\<\ul{\eta}_\xi,\1b\>
\eeqa
and hence $\ul{\eta}_\xi$ is euclidean invariant.
Let $e\in S^{d-1}$ be a euclidean time direction and let 
$\1b\in\ul{\6B}(e)$ be localized in the half space 
$\7R_+e+e^\perp$. Then we obtain from the reflexion positivity 
of $\eta$ that 
\beqa
\<\eta, j_e(\1b(\lam))\1b(\lam)\>&\geq&  0
\eeqa
since $\1b(\lam)\in\6B(e)$ is localized in the half space 
$\7R_+e+e^\perp$ for each $\lam$.
Since $\xi$ is a positive functional, we get  
\beqa
\<\ul{\eta}_\zeta, j_e(\1b)\1b\>&=&
\int \8d\xi(\lam) \ \<\eta,j_e(\1b(\lam))\1b(\lam)\>
\ \ \geq \ \ 0 
\eeqa
and the reflexion positivity for $\ul{\eta}_\xi$ follows.

Finally, regularity holds true for $\eta$ since 
$\ul{\be}$ is strongly continuous. Namely, for 
$\1b_1,\1b_2,\1b_3\in\ul{\6B}$ we have
\beqa
\lim_{h\to 1}|\<\ul{\eta}_\xi,\1b_1[\ul{\be}_h(\1b_2)-\1b_2]\1b_3\>|
&\leq&
\|\1b_1\| \ \|\1b_3\| \ \lim_{h\to 1} \|\ul{\be}_h(\1b_2)-\1b_2\|
\vs\vs
&=&0
\eeqa
and the regularity follows.
\epr

\paragraph{\em Remark.}
Taking scaling limits by making use of limit 
functionals is slightly more general than the method 
of taking subnets as it has been used by D. Buchholz and 
R. Verch. We make some more detailed comments 
on this fact 
in Appendix \ref{app1}.
\skb 

\paragraph{\em Quantum fields, constructed from 
the euclidean scaling limits.}
According to Proposition \ref{pro1} the triple 
$(\ul{\6B},\ul{\be},\ul{\eta}_\zeta)$ is an euclidean field. 
We recall here the procedure of Step 1 and Step 2, 
given in the previous section,
in order to fix our notations. 

\paragraph{\em Step 1':}
A Hilbert space $\2H_\zeta$ and a linear map
\beqa
\Psi_\zeta:\ul{\6B}(e)\to \2H_\zeta
\eeqa
can be constructed from $(\ul{\6B},\ul{\be},\ul{\eta}_\zeta)$, 
where $\Psi_\zeta$ is uniquely determined by 
\beqa
\<\Psi_\zeta[\1b_1],\Psi_\zeta[\1b_2]\>&=&\<\ul{\eta}_\zeta,j_e(\1b_1)\1b_2\>
\eeqa
for each $\1b_1,\1b_2\in\ul{\6B}(e)$. We obtain a 
unitary strongly continuous representation $U_\zeta$ of the 
Poincar\'e group on $\2H_\zeta$ by \cite{FrohOstSeil}, 
with invariant vector $\Om_\zeta=\Psi_\zeta[\11]$.

\paragraph{\em Step 2':}
As already mentioned, the strong continuity of the 
group homomorphism $\ul{\be}$ might cause the 
problem that all operators 
in $\ul{\6B}$ which are localized in the time-slice $e^\perp$
are multiples of the identity. Thus we wish to
find an extension $(\hat{\ul{\6B}},\hat{\ul{\be}},\hat{\ul{\eta}}_\zeta)$
of the euclidean net $(\ul{\6B},\ul{\be},\ul{\eta}_\zeta)$
such that non trivial time slice operators may be found in there.

As we are going to show in Appendix \ref{app2} (Lemma \ref{lemext0}),
there indeed exists an extension  
$(\hat{\ul{\6B}},\hat{\ul{\be}},\hat{\ul{\eta}}_\zeta)$
of the euclidean field $(\ul{\6B},\ul{\be},\ul{\eta}_\zeta)$
with the following property: Let $\6J_\zeta$ be the 
ideal which is annihilated by the GNS representation 
of $\hat{\ul{\eta}}_\zeta$. Then the algebra 
$\hat{\ul{\6B}}/\6J_\zeta$ is the extension 
of $\ul{\6B}/\6J_\zeta$ which can be obatained by the 
completion procedure in Step 2 of the previeous section.
In particular, the algebra $\hat{\ul{\6B}}$ is
independent of the limit functional $\zeta$.

Due to Proposition \ref{pro01}, for an operators
$\1b\in\ul{\6B}(e,s)$, localized near the time-slice $e^\perp$,
a bounded operator $\pi_{(s,\zeta)}(\1b)$
on $\2H_\zeta$ is given by 
\beqa
\pi_{(s,\zeta)}(\1b)\Psi_\zeta[\1b_1]&:=&\Psi_\zeta[\1b\ul{\be}_{se}\1b_1] 
\eeqa
and by Proposition \ref{pro02}, for a time slice operator 
$\1b_0\in\hat{\ul{\6B}}(e^\perp)$, a bounded operator 
$\pi_\zeta(\1b_0)$ on $\2H_\zeta$ is defined by 
\beqa
\pi_{\zeta}(\1b_0)\Psi_\zeta[\1b_1]&:=&\Psi_\zeta[\1b_0\1b_1] \ \ .
\eeqa
Thus we can build a Haag-Kastler net:
For a double cone $\9O$, we define $\6A_\zeta(\9O)$ to be the 
C*-subalgebra  in $\6B(\2H_\zeta)$ which is generated by 
all operators 
\beqa
\Pi_\zeta[f,\1b] &=& \int \8dg \ f(g) \ U_\zeta(g)\pi_\zeta(\1b) U_\zeta(g)^*
\eeqa
with $\1b\in\hat{\ul{\6B}}(\9V)$, $f\in\9C^\infty_0(\Poin)$, such that  
$g\9V\subset\9O$ for each 
Poincar\'e transformation $g$ in the support of $f$.
The prescription 
\beqa
\9O\mapsto\6A_\zeta(\9O)
\eeqa 
is an isotonous net of C*-algebras and by putting 
$\al_{(\zeta,g)}:=\Ad(U_\zeta(g))$, for each $g\in\Poin$, and 
$\om_\zeta:=\<\Om_\zeta,(\cdot)\Om_\zeta\>$ 
we obtain the quantum field $(\6A_\zeta,\al_\zeta,\om_\zeta)$.

\paragraph{\em Scaling limits for the constructed 
Haag-Kastler net.}
The scaling algebra $\ul{\6A}$ can be expressed in   
terms of the time-zero algebras $\hat{\ul{\6B}}(\9V)$
of the extended net, introduced in previous paragraph.
Each operator $\1b$ can be identified with a map 
$\1b:\7R_+\to\hat{\6B}(e^\perp)$ (Lemma \ref{lemext1}) 
which assigns to each scaling parameter a time slice operator in 
$\hat{\6B}(e^\perp)$. The scaling algebra 
$\ul{\6A}$ is generated by all functions 
\beqa
\ul\Pi[f,\1b]:\lam&\mapsto&\ul{\Pi}[f,\1b](\lam) \ := \ 
\int \8dg \ f(g) \ U_\lam(g)\pi(\1b(\lam))U_\lam(g)^*
\eeqa
where $f\in\9C^\infty_0(\Poin)$ is a smooth function on the 
Poincar\'e group with compact support and 
$\1b\in\hat{\ul{\6B}}(e^\perp)$ is a 
time slice operator.
Here, $U_\lam$ is the scaled representation of the 
Poincar\'e group on $\2H$, which is defined according to 
$U_\lam(g):=U(\lam\circ g\circ\lam^{-1})$ for each 
Poincar\'e transformation $g$.
 
As also described in the introduction, a group homomorphism $\ul{\al}$ from
the Poincar\'e group into the automorphism group 
of the scaling algebra $\ul{\6A}$, acting covariantly on the net 
$\9O\mapsto\ul{\6A}(\9O)$, is simply defined according to
\beqa
(\ul{\al}_g\1a)(\lam)&:=&\al_{\lam\circ g\circ\lam^{-1}}\1a(\lam)
\eeqa
for each $\1a\in\ul{\6A}$. 
 
Of course, we have to ensure that the 
algebra $\ul{\6A}$ contains renormalization group orbits 
with the correct scaling property in 
configuration space as well as in 
momentum space:

\bpro\label{proscaling}
The group homomorphism $\ul{\al}$ is strongly continuous 
on $\ul{\6A}$:
\beqa
\lim_{g\to 1}\|\1a-\ul{\al}_g\1a\|&=&0
\eeqa
for each $\1a\in\ul{\6A}$.
\epro
\bpr
We first check the continuety for all generators
$\ul\Pi[f,\1b]$, 
where $f\in\9C^\infty_0(\Poin)$ is a smooth function on the 
Poincar\'e group with compact support and 
$\1b\in\hat{\ul{\6B}}(e^\perp)$ is a 
time slice operator:
\beqa
&&\lim_{g\to 1}\|\ul\Pi[f,\1b]-\ul{\al}_g\ul\Pi[f,\1b]\|
\vs\vs
&=&
\lim_{g\to 1}\|\ul\Pi[f,\1b]-\ul\Pi[f\circ g^{-1},\1b]\|
\vs\vs
&=&\lim_{g\to 1}\|\ul\Pi[f-f\circ g^{-1},\1b] \|
\vs\vs
&\leq&
\sup_{\lam\in\7R_+}\|\1b(\lam)\| \ 
\lim_{g\to 1} \int \8dg' \ |f(g')-f(g^{-1}g')|
\vs\vs
&=&0 \ \ .
\eeqa
Now let $\1a_1,\1a_2$ be two operators with 
$\lim_{g\to 1}\|\1a_j-\ul{\al}_g\1a_j\|=0$ for $j=1,2$ then we obtain
\beqa
&&\lim_{g\to 1}\|\1a_1\1a_2-\ul{\al}_g(\1a_1\1a_2)\|
\vs\vs
&\leq&\|\1a_1\| \ \lim_{g\to 1}\|\1a_2-\ul{\al}_g(\1a_2)\|
+ \|\1a_2\| \ \lim_{g\to 1}\|\1a_1-\ul{\al}_g(\1a_1)\|
\vs\vs
&=&0
\eeqa
which implies that strong continuety is valid 
for all finite linear combinations of products of generators.
For each operator $\1a\in\ul{\6A}$ and for each $\eps>0$ 
we can find a 
finite linear combination of products of generators $\1a_\eps$ 
such that $\|\1a-\1a_\eps\|<\eps/2$.
Thus we get
\beqa
\lim_{g\to 1}\|\1a-\ul{\al}_g(\1a)\|&\leq&
2\|\1a-\1a_\eps\|+ \lim_{g\to 1}\|\1a_\eps-\ul{\al}_g(\1a_\eps)\|
\vs\vs
&<&\eps
\eeqa
and hence $\lim_{g\to 1}\|\1a-\ul{\al}_g(\1a)\|=0$ 
for all $\1a\in\ul{\6A}$.
\epr

For a given limit functional $\zeta$ we build the 
scaling limit $(\ul{\6A}_\zeta,\ul{\al}_\zeta,\ul{\om}_\zeta)$
of the quantum field $(\6A,\al,\om)$
in the following manner:
A vacuum state $\ul{\om}_\zeta$ is given 
according to the prescription 
\beqa
\<\ul{\om}_\zeta,\1a\>&:=&\int \8d\zeta(\lam) \  \<\om,\1a(\lam)\>
\eeqa
for each $\1a\in\ul{\6A}$. Let $\6J_\zeta$ be the two sided 
ideal in $\ul{\6A}$ which is annihilated by the GNS 
representation of $\ul{\om}_\zeta$. We build the 
quotient C*-algebra 
\beqa
\ul{\6A}_\zeta&:=&\ul{\6A}/\6J_\zeta
\eeqa
and we denote by $\1q_\zeta$ the corresponding canonical projection
onto the quotient. The group homomorphism $\ul{\al}$
can be lifted to a group homomorphism $\ul{\al}_\zeta$ from
the Poincar\'e group into the automorphism group 
of $\ul{\6A}_\zeta$ by
\beqa
\ul{\al}_{(\zeta,g)}\circ\1q_\zeta&=&\1q_\zeta\circ\ul{\al}_g
\eeqa
for each Poincar\'e trnsformation $g$.
As a consequence the assignment 
\beqa
\9O&\mapsto&\ul{\6A}_\zeta(\9O) \ \ := \ \ \1q_\zeta(\ul{\6A}(\9O))
\eeqa
together with the group homomorphism $\ul{\al}_\zeta$ is a Poincar\'e 
covariant Haag-Kastler net. 
 

\paragraph{\em The main result.}
We are now prepared to show that the 
subsequent two step procedures applied to a 
given euclidean field $(\6B,\be,\eta)$, lead to equivalent results.

\bdes

\itno 1
For a given limit functional $\zeta\in\6S[\8{C}(\7R_+)]$ we first build  
the euclidean field 
$(\ul{\6B},\ul{\be},\ul{\eta}_\zeta)$ and then the corresponding 
quantum field theory model 
$(\6A_\zeta,\al_\zeta,\om_\zeta)$.

\itno 2
On the other hand, we first build the quantum field theory model
$(\6A,\al,\om)$ and then we construct the 
scaling limit  
$(\ul{\6A}_\zeta,\ul{\al}_\zeta,\ul{\om}_\zeta)$
with respect to a limit point $\zeta\in\6S[\8{C}(\7R_+)]$.

\edes

\bthe\label{the1}
For each limit functional $\zeta\in\6S[\8{C}(\7R_+)]$, 
the quantum fields $(\6A_\zeta,\al_\zeta,\om_\zeta)$
and $(\ul{\6A}_\zeta,\ul{\al}_\zeta,\ul{\om}_\zeta)$
are equivalent.
\ethe

The complete proof of Theorem \ref{the1} is given in the Appendix \ref{app2}
and we only describe the  main idea of it here. 

\paragraph{\em Sketch of the proof:}
In the first step, a Poincar\'e covariant representation $\ul{\pi}_\zeta$ of 
the scaling algebra $\ul{\6A}$ on a Hilbert space 
$\ul{\2H}_\zeta$ is constructed and it can be 
shown that $\ul{\pi}_\zeta$ is equivalent to the 
GNS representation of the scaling limit state $\ul{\om}_\zeta$.
In the second step, an isometry $\1u_\zeta$ from the 
Hilbert space $\2H_\zeta$ to $\ul{\2H}_\zeta$ is constructed such that 
\beq\label{iso}
\1u_\zeta\Pi_\zeta[f,\1b]\1u_\zeta^*&=&
\ul{\pi}_\zeta(\ul{\Pi}[f,\1b])
\eeq
holds true for each smooth function with compact support 
$f$ on the Poincar\'e group and for each time zero 
operator $\1b$ in the extended scaling algebra $\hat{\ul{\6B}}(e^\perp)$.
Note, that the operator $\Pi_\zeta[f,\1b]$ is contained in 
the C*-algebra $\6A_\zeta$, corresponding to procedure {\it (1)},
whereas $\ul{\pi}_\zeta(\ul{\Pi}[f,\1b])$ can be identified with 
an operator in the C*-algebra $\ul{\6A}_\zeta$ 
corresponding to procedure {\it (2)}. As a consequence the
map
\beqa
\iota_\zeta:\1a &\mapsto&\1u_\zeta\1a\1u_\zeta^*
\eeqa
yields an isomorphism of the algebras $\6A_\zeta$ and 
$\ul{\6A}_\zeta$ which respects the net structure, 
$\iota_\zeta(\6A_\zeta(\9O))=\ul{\6A}_\zeta(\9O)$ for each 
double cone $\9O$. Moreover, $\iota_\zeta$ intertwines the 
automorphisms $\al_\zeta$ and $\ul{\al}_\zeta$, 
i.e. $\iota_\zeta\circ\al_\zeta=\ul{\al}_\zeta\circ\iota_\zeta$,
and maps the vacuum state $\ul{\om}_\zeta$ to 
$\om_\zeta$, i.e. $\ul{\om}_\zeta\circ\iota_\zeta=\om_\zeta$.
 
The crucial point in order to prove Equation (\ref{iso}) 
is to establish the fact that $\1u_\zeta$ 
intertwines the corresponding representations 
of the Poincar\'e group. More precisely, the Hilbert space 
$\2H_\zeta$ (respectively $\ul{\2H}_\zeta$), carry a 
strongly continuous unitary representation 
$U_\zeta$ (respectively $\ul{U}_\zeta$) of the Poincar\'e group and 
one has to show that 
\beqa
\1u_\zeta U_\zeta(g)\1u_\zeta^*&=&\ul{U}_\zeta(g)
\eeqa
holds true for each Poincar\'e transformation $g$.
This intertwiner property can be verified as follows:
Let $B$ be the generator of a Poincar\'e transformation. 
Then there is a dense subspace $\2D\subset\ul{\2H}_\zeta$
such that for each pair $\psi_1,\psi_2$ the functions
\beqa
&&\ul{F}:z\mapsto \<\psi_1, \ul{U}_\zeta(\exp(zB))\psi_2\>
\vs\vs
&&F:z\mapsto \<\1u_\zeta^*\psi_1, U_\zeta(\exp(zB))\1u_\zeta^*\psi_2\>
\eeqa
are holomorphic within a strip $\7R+\8iI$, where $I$ is some 
connected open subset in $\7R$. Within the pure imaginary points,
both functions can be expressed explicitly in terms of euclidean 
correlation functions within the scaling limit state 
$\ul{\eta}_\zeta$ and one finds that
\beqa
\ul{F}(\8i\tau)&=&F(\8i\tau)
\eeqa
is valid for each $\tau\in I$. Hence one concludes 
$F=\ul{F}$ and the interwiner property follows since 
$\2D$ is dense in $\ul{\2H}_\zeta$.

\section{Concluding remarks}
\label{sc5} 
We have proven that 
the quantum field theories, which can be reconstructed from 
euclidean scaling limit theories, are scaling limit 
theories of the quantum field theory which can be reconstructed from
the underlying euclidean field theory. 

This fact leads us into a comfortable position which  
can be also motivated by the consideration of euclidean field theories
with cutoffs. Usually, to a regularized euclidean model  
the construction scheme \cite{Schl97} cannot even by applied. 
On the other hand, the euclidean counterpart 
of the analysis by D. Buchholz and R. Verch \cite{BuVer95,BuVer97} can be
formulated for euclidean field theory models in the C*-setting 
as we have discussed during this paper. The procedure, given in 
Section \ref{sc2}, is quite general and it 
still makes sense also for regularized euclidean models
with a infra-red and an ultra-violet cutoff.

We expect that scaling limit theories of euclidean field theories
within a finite volume are 
essentially independent of the volume cutoff.
We propose to regard 
a finite volume euclidean field theory in $d$ dimensions as a 
field theory on the scaled $d$-sphere $rS^d\subset\7R^{d+1}$, where 
$r$ is the volume cutoff. The corresponding euclidean net $\6B$ 
carries a covariant action of the $d+1$-dimensional rotation group 
$\8O(d+1)$ and the functional $\eta$ is invariant under this action. 
For a point $x_0\in rS^d$, the stabilizer subgroup of $x_0$ in 
$\8O(d+1)$ is isomorphic to $\8O(d)$. If the scaling limit procedure 
is performed at $x_0$, the invariance under the stabilizer 
subgroup should remain as a  $\8O(d)$ invariance within the scaling 
limit. The translation invariance should then enter from the 
fact that $\eta$ is invariant under the full group $\8O(d+1)$.
Note, that $r$ is replaced by $r_\lam=\lam^{-1}r$ for the 
scaled theory. Hence one expects that the scaling limit theories
are models within an infinite volume. 

Keeping in mind that the minkowskian analogue of the euclidean 
$d$-sphere $rS^d\subset\7R^{d+1}$ is the de Sitter space, 
it should be possible, by exploring the analytic structure of de Sitter space, 
to construct from a given euclidean field theory 
$(\6B,\al,\eta)$ on the sphere $rS^d$ 
a quantum field theory $(\6A,\al,\om)$
in de Sitter space. 
According to the considerations of H. J. Borchers 
and D. Buchholz \cite{BorBu98} 
we conjecture that the reconstructed state $\om$
fulfills the so called {\em geodesic KMS condition}, i.e.
for any geodesic observer the state $\om$ looks like 
an equilibrium state.

\subsubsection*{{\it Acknowledgment:}}
I am grateful to Prof. Jakob Yngvason for 
supporting this investigation with hints and many ideas.
This investigation is financially supported by the 
{\em Jubil\"aumsfond der Oesterreichischen Nationalbank} 
which is also gratefully acknowledged.
Finally I would like to thank the 
Erwin Schr\"odinger International Institute for Mathematical Physics, 
Vienna (ESI) for its hospitality.
\newpage

\begin{appendix}
\section{Remarks on limit functionals}
\label{app1}
As a tool for taking scaling limits we have used 
the limit functional on the 
corona algbera $\8{C}(\7R_+)=\9F_\8b(\7R_+)/\9F_0(\7R_+)$.
More general, we replace $\7R_+$ by any partially ordered directed 
set $\Lam$, and we consider the C*-algebra $\9F_\8b(\Lam)$ of all bounded 
functions on $\Lam$ and the closed two-sided ideal $\9F_0(\Lam)$
in $\9F_\8b(\Lam)$
which is generated by functions $f\in\9F_\8b(\Lam)$ 
with $\lim_{\lam\in\Lam}f(\lam)=0$. The states $\zeta\in\6S[\8C(\Lam)]$
on the quotient C*-algebra 
\beqa
\8C(\Lam)&=&\9F_\8b(\Lam)/\9F_0(\Lam)
\eeqa
are called the {\em limit functionals} with respect to $\Lam$.

Let $V$ be a locally convex linear space whose topology 
is induced by a family of semi norms $\1p\in\9P$.
A function $\1w:\Lam\to V'$ from the partially ordered set 
$\Lam$ to the dual space $V'$ is called {\em bounded} if there exists a
semi norm $\1p\in\9P$ and a constant $K>0$ such that  
\beqa
\sup_{\lam\in\Lam}|\<\1w(\lam),v\>| &\leq& K \ \1p(v) 
\eeqa  
holds true for each $v\in V$. 
For each limit functional $\zeta\in \6S[\8C(\Lam)]$ we obtain a
linear functional $\1w_\zeta$ on $V$ by the prescription 
\beqa
\<\1w_\zeta,v\>&:=&\int\8d\zeta(\lam) \ \<\1w(\lam),v\> \ \ .
\eeqa
The functional $\1w_\zeta$ is well defined since for each 
$v\in V$ the function $[\lam\mapsto \<\1w(\lam),v\>]$ is 
contained in $\9F_\8b(\Lam)$. In particular, $\1w_\zeta$ fulfills 
for each $v\in V$ the 
estimate 
\beqa
|\<\1w_\zeta,v\>| &\leq& K \ \1p(v) 
\eeqa
and $\1w_\zeta$ is a continuous functional on $V$.

As a special case, let $V$ be a Banach space with norm $\|\cdot\|$
and let $\|\cdot\|'$ be the dual norm on $V'$.
For each bounded function $\1w:\Lam\to V'$, 
we create new elements in $V'$ in two different ways:
\bdes
\itno 1
According to the Banach-Alaoglu theorem, the 
set $\{\1w(\lam),\lam\in\Lam\}$ is precompact and 
there exists a partially ordered directed set $J$ and a subnet  
$\iota:J\to\Lam$ such that 
\beqa
\<\1w_\iota,v\>&:=&\lim_{j\in J}\<\1w(\iota(j)),v\>
\eeqa
for each $v\in V$.

\itno 2
We also can choose a limit functional 
$\zeta\in \6S[\8C(\Lam)]$ and build the linear functional 
$\1w_\zeta\in V'$ as described above.
\edes

Choosing the state $\1w_\iota$, via a subnet $\iota:J\to\Lam$,
is essentially the same as choosing a limit functional
$\zeta\in \6S[\8C(\Lam)]$. More precisely:

\bpro
For each subnet $\iota:J\to\Lam$ for which the weak*- limit 
$\1w_\iota=w^*-\lim_{j\in J}\1w(\iota(j))$ exists in $V'$, there 
exists a limit functional $\zeta\in\6S[\8C(\Lam)]$ such 
that 
\beqa
\1w_\iota&=&\1w_\zeta \ \ .
\eeqa 
\epro
\bpr
The subnet $\iota:J\to\Lam$ induces a *-homomorphism 
$\iota^*:\9F_\8b(\Lam)\to\9F_\8b(J)$ by 
$\iota^*f=f\circ\iota$. One observes that for 
each state $\hat\zeta\in\6S[\8C(J)]$ on the corona algebra of $J$,
the state $\zeta=\hat\zeta\circ\iota^*$ yields a 
state on $\8C(\Lam)$. Namely, let $f\in \9F_0(\Lam)$ 
then $\lim_{\lam\in\Lam}f(\lam)=0$ and thus 
$\lim_{j\in J}f(\iota(j))=0$ for each subnet $\iota:J\to\Lam$.
Hence $\iota^*\9F_0(\Lam)\subset \9F_0(J)$ and the 
state $\zeta=\hat\zeta\circ\iota^*$ annihilates $\9F_0(\Lam)$
which implies $\zeta\in \6S[\8C(\Lam)]$.
Since for each $v\in V$ the function 
$[j\mapsto\<\1w(\iota(j))-\1w_\iota,v\>]$ is contained in 
$\9F_0(J)$ we conclude 
\beqa
\<\1w_\iota,v\>&=&\int\8d\hat\zeta(j) \ \<\1w(\iota(j)),v\>
\vs\vs
&=&
\int\8d (\hat\zeta\circ\iota^*)(\lam) \ \<\1w(\lam),v\> 
\vs\vs
&=&\<\1w_\zeta,v\>
\eeqa
which implies the proposition.
\epr

\section{Notes on scaling limits of euclidean fields}
\label{app10}
In order to construct from the euclidean data 
$(\ul{\6B},\ul{\be},\ul{\eta}_\zeta)$ a quantum field theory
model, we discuss now an appropriate extension 
of the scaling algebra $\ul{\6B}$.

Let $(\2K,\tau,E)$ be the GNS triple of $\eta$. According to the 
euclidean invariance of $\eta$, there exists a strongly continuous 
representation $V$ of the euclidean group on $\2K$ which is
uniquely determined by
\beqa
V(h)\tau(b)E&=&\tau(\be_hb)E
\eeqa
for each $b\in\6B$ and for each euclidean transformation $h$.
Let $\ul{\2K}_o$ be the linear space, spanned by bounded functions
$\ul{\psi}:\7R_+\to\2K$ of the form
\beqa
\ul{\psi}_\lam&=& \int \8dh \ f(h) \ V_\lam(h)\ul{\psi}^0_\lam
\eeqa
where $\ul{\psi}^0:\7R_+\to\2K$ is any bounded function 
and $V_\lam(h)=V(\lam\circ h\circ \lam^{-1})$ is the scaled representation.
We introduce a locally convex topology on $\ul{\2K}_o$ which is 
induced by semi norms $\|\cdot\|_\xi$, where 
$\xi$ is a state on $\6F_\8b(\7R_+)$:
\beqa
\|\ul{\psi}\|_\xi^2&:=&
\int\8d\xi(\lam) \ \<\ul{\psi}_\lam,\ul{\psi}_\lam\> \ \ .
\eeqa
The closure of $\ul{\2K}_o$ with respect to this topology is 
denoted by $\ul{\2K}$.  The prescription
\beqa
[\ul{\tau}(\1b)\ul{\psi}]_\lam&:=&\tau(\1b(\lam))\ul{\psi}_\lam
\eeqa
yields a faithful representation of the scaling algebra 
on $\ul{\2K}$ by bounded operators $\ul{\tau}(\1b)$.

We also introduce semi norms on $\ul{\6B}$, namely for a 
state $\xi$ on  $\6F_\8b(\7R_+)$ and for a vector
$\ul{\Psi}$ we introduce the semi norm 
$\|\cdot\|_{(\xi,\ul{\psi})}$ by
\beqa
\|\1b\|_{(\xi,\ul{\psi})}&=&\|\ul{\tau}(\1b)\ul{\psi}\|_\xi
\eeqa
and the closure of $\ul{\6B}$ within this topology is 
denoted by $\hat{\ul{\6B}}$. Note that the 
group homomorphism $\ul{\be}$ can be extended to a 
group homomorphism $\hat{\ul{\be}}$ form
the euclidean group into the automorphism group of $\hat{\ul{\6B}}$.

For each state $\xi$ on $\9F_\8b(\7R_+)$ there is a Hilbert space 
$\ul{\2K}_\xi$  and a linear map 
$\1p_\xi:\ul{\2K}\to\ul{\2K}_\xi$ which is uniquely determined by
\beqa
\<\1p_\xi[\ul{\psi}_1],\1p_\xi[\ul{\psi}_2]\>
&=&\int\8d\xi(\lam) \ \<\ul{\psi}_{1,\lam},\ul{\psi}_{2,\lam}\> \ \ .
\eeqa
We also use, which is sometimes convenient, the suggestive notation 
\beqa
\1p_\xi[\ul{\psi}]&=&\int^\oplus \8d\xi(\lam) \ \ul{\psi}_\lam \ \ .
\eeqa

\blem\label{lemext0}
For each state $\xi$ on $\9F_\8b(\7R_+)$ the following statements are valid:
\bdes
\itno 1 
There exists 
a strongly continuous unitary representation 
$\ul{V}_\xi$ on $\ul{\2K}_\xi$ of the euclidean group.

\itno 2
There exists a *-representation $\ul{\tau}_\xi$ on $\ul{\2K}_\xi$ 
of the extended scaling algebra $\hat{\ul{\6B}}$ 
and a vector $\ul{E}_\xi\in\ul{\2K}_\xi$ such that
\beqa
\<\ul{\eta}_\xi,\1b\>&=&\<\ul{E}_\xi,\ul{\tau}_\xi(\1b)\ul{E}_\xi\>
\eeqa
is valid for each $\1b\in\ul{\6B}$. 

\itno 3
For each $h\in\8E(d)$ and for each $\1b\in\hat{\ul{\6B}}$ the identity  
\beqa
\ul{\tau}_\xi(\hat{\ul{\be}}_h\1b)&=&
\ul{V}_\xi(h)\ul{\tau}_\xi(\1b)\ul{V}_\xi(h)^*
\eeqa
holds true.
\edes
\elem
\bpr
We define the representation $\ul{V}_\xi$ of the euclidean group by 
\beqa
\ul{V}_\xi(h)\1p_\xi[\ul{\psi}]&=&\int^\oplus \8d\xi(\lam) \
V_\lam(h) \ \ul{\psi}_\lam 
\eeqa
for each $h\in\8E(d)$ and for each 
$\ul{\psi}\in\ul{\2K}$. 
For each $\1b\in\ul{\6B}$ and for each $\ul{\psi}\in\ul{\2K}$ 
the map $\1b\mapsto \ul{\tau}(\1b)\ul{\psi}$ is continuous 
as a linear function form $\hat{\ul{\6B}}$ to $\ul{\2K}$ since 
for each $\xi$ we have 
\beqa
\|\ul{\tau}(\1b)\ul{\psi}\|_\xi&=&\|\1b\|_{(\xi,\ul{\psi})}
\eeqa
and $\ul{\tau}$ can uniquely be extended to 
representation $\hat{\ul{\tau}}$ of $\hat{\ul{\6B}}$ 
on $\ul{\2K}$. Now we define the representation 
$\ul{\tau}_\xi$ by 
\beqa
\ul{\tau}_\xi(\1b)\1p_\xi[\ul{\psi}]&:=&\1p_\xi[\hat{\ul{\tau}}(\1b)\ul{\psi}]
\eeqa
for each $\1b\in\hat{\ul{\6B}}$ and for each 
$\ul{\psi}\in\ul{\2K}$.

\bdes
\itno 1
We first show that $\ul{V}_\xi$ is indeed a strongly continuous 
representation of the euclidean group.
The vectors of $\ul{\psi}_\xi$ of the form 
\beqa
\ul{\psi}_\xi&=& \int^\oplus \8d\xi(\lam) \ 
\int \8dh \ f(h) \ V_\lam(h)\ul{\psi}^0_\lam \ \ ,
\eeqa
where $\ul{\psi}^0:\7R_+\to\2K$ is any bounded function,
span a dense subspace in $\ul{\2K}_\xi$. We compute 
\beqa
\ul{V}_\xi(h)\ul{\psi}_\xi &=&
\int^\oplus \8d\xi(\lam) \ 
\int \8dh' \ f(h') \ V_\lam(hh')\ul{\psi}^0_\lam
\vs\vs
&=&
\int^\oplus \8d\xi(\lam) \ 
\int \8dh' \ f(h^{-1}h') \ V_\lam(h')\ul{\psi}^0_\lam 
\eeqa
which yields 
\beqa
\|\ul{V}_\xi(h)\ul{\psi}_\xi-\ul{\psi}_\xi\|
&\leq& 
\int \8dh' \ |f(h^{-1}h')-f(h')| \ \sup_{\lam\in\7R_+}\|\ul{\psi}^0_\lam\|
\eeqa
and we conclude that 
\beqa
\lim_{h\to 1}\|\ul{V}_\xi(h)\ul{\psi}_\xi-\ul{\psi}_\xi\|&=&0
\eeqa
holds true for all $\ul{\psi}_\xi$ in a dense subspace of $\ul{\2K}_\xi$
and the statement {\it (1)} follows.

\itno 2
The Hilbert space $\ul{\2K}_\xi$ 
contains a distinguished vector $\ul{E}_\xi$ 
\beqa
\ul{E}_\xi&=&\int^\oplus \8d\xi(\lam) \ E
\eeqa
which is invariant under the representation $\ul{V}_\xi$. 
This vector yields a state $\hat{\ul{\eta}}_\xi$ on the extended 
scaling algebra $\hat{\ul{\6B}}$ by
\beqa
\<\hat{\ul{\eta}}_\xi,\1b\>&:=&\<\ul{E}_\xi,\ul{\tau}_\xi(\1b)\ul{E}_\xi\>
\eeqa
for $\1b\in\hat{\ul{\6B}}$. According to the construction 
of $\ul{\2K}_\xi$ we easily observe that $\hat{\ul{\eta}}_\xi$
is an extension of $\ul{\eta}_\xi$:
\beqa
\<\ul{E}_\xi,\ul{\tau}_\xi(\1b)\ul{E}_\xi\>
&=&
\int\8d\xi(\lam) \ \< E, \tau(\1b(\lam))E\>
\vs\vs
&=&
\int\8d\xi(\lam) \ \<\eta,\1b(\lam)\>
\vs\vs
&=&
\<\ul{\eta}_\xi,\1b\>
\eeqa
for each $\1b\in\ul{\6B}$ which implies  {\it (2)}.

\itno 3
Statement {\it (3)} can easily be verfied, from the definition of 
the representations $\ul{\tau}_\xi$ and $\ul{V}_\xi$.

\edes
\epr

The extended state $\hat{\ul{\eta}}_\xi$ is euclidean invariant, 
reflexion positive, and regular. 
In particular, the regularity follows from the fact that 
$\ul{V}_\xi$ is a strongly continuous representation of the 
euclidean group.
Hence the triple $(\hat{\ul{\6B}},\hat{\ul{\be}},\hat{\ul{\eta}}_\xi)$
is a euclidean field which extends $(\ul{\6B},\ul{\be},\ul{\eta}_\xi)$.

\blem\label{lemext1}
The time zero algebra $\hat{\ul{\6B}}(e^\perp)$ can be identified
with the C*-algebra generated by bounded 
functions $\1b:\lam\mapsto\1b(\lam)\in\hat\6B(\lam\9V)$
mapping each $\lam\in\7R_+$ into the algebra $\hat\6B(\lam\9V)$ 
for some convex and bounded region $\9V\subset e^\perp$.
\elem
\bpr
For each $\lam\in\7R_+$ the representation $\ul{\tau}_\lam$ is a
representation of $\hat{\ul{\6B}}$ on $\2K$ since we have for each 
$\1b\in\ul{\6B}$
\beqa
\ul{\tau}_\lam(\1b)\1p_\lam[\ul{\psi}]&=&\tau(\1b(\lam))\ul{\psi}_\lam \ \ .
\eeqa
We have assumed that $\tau$ is a faithful representation and
we may define for each time zero operator 
$\1b\in\hat{\ul{\6B}}(\9V)$, $\9V\subset e^\perp$, the operator
\beqa
\1b(\lam)&:=&\tau^{-1}(\ul{\tau}_\lam(\1b))
\eeqa
which is contained in $\hat{\6B}$. According to the localizing property 
of $\1b$ we conclude
\beqa
\1b(\lam)&\in&\bigcap_{s\in\7R_+}\hat{\6B}([0,\lam s)e\times\lam\9V)
\ \ = \ \ \hat{\6B}(\lam\9V) \ \ .
\eeqa
Since the W*-algebra $\hat{\6B}$ is closed in the strong operator topology, 
for a given function $\1b:\lam\mapsto\1b(\lam)\in\hat\6B(\lam\9V)$
there is a net of operators $(\1b_j)_{j\in J}$ in 
$\ul{\6B}$, converging in 
$\hat{\ul{\6B}}$, such that 
\beqa
\lim_{j\in J}\|\1b_j-\1b\|_{(\lam,\ul{\psi})}
&=&
\lim_{j\in J}\|\tau(\1b_j(\lam)-\1b(\lam))\ul{\psi}_\lam\| \ \ = \ \ 0
\eeqa
for each $\ul{\psi}\in\ul{\2K}$, pointwise in $\lam\in\7R_+$.
\epr

\section{Notes on scaling limits of quantum fields constructed form
euclidean data}
\label{app20}
Let $\ul{\2H}_o$ be the linear space, spanned by bounded functions
$\ul{\psi}:\7R_+\to\2H$ of the form
\beqa
\ul{\psi}_\lam&=& \int \8dg \ f(g) \ U_\lam(g)\ul{\psi}^0_\lam
\eeqa
where $\ul{\psi}^0:\7R_+\to\2H$ is any bounded function 
and $U_\lam(g)=U(\lam\circ g\circ \lam^{-1})$ is the scaled representation.
We introduce a locally convex topology on $\ul{\2H}_o$ which is 
induced by semi norms $\|\cdot\|_\xi$, where 
$\xi$ is a state on $\6F_\8b(\7R_+)$:
\beqa
\|\ul{\psi}\|_\xi^2&:=&
\int\8d\xi(\lam) \ \<\ul{\psi}_\lam,\ul{\psi}_\lam\> \ \ .
\eeqa
The closure of $\ul{\2H}_o$ with respect to this topology is 
denoted by $\ul{\2H}$.  The prescription
\beqa
[\ul{\pi}(\1a)\ul{\psi}]_\lam&:=&\1a(\lam)\ul{\psi}_\lam
\eeqa
yields a faithful representation $\ul{\pi}$ of the scaling algebra $\ul{\6A}$ 
on $\ul{\2H}$ by bounded operators.

As in the euclidean case, for 
each state $\xi$ on $\9F_\8b(\7R_+)$ there is a Hilbert space 
$\ul{\2H}_\xi$  and a linear map 
$\1q_\xi:\ul{\2H}\to\ul{\2H}_\xi$ which is uniquely determined by
\beqa
\<\1q_\xi[\ul{\psi}_1],\1q_\xi[\ul{\psi}_2]\>
&=&\int\8d\xi(\lam) \ \<\ul{\psi}_{1,\lam},\ul{\psi}_{2,\lam}\> 
\eeqa
and we also write
\beqa
\1q_\xi[\ul{\psi}]&=&\int^\oplus \8d\xi(\lam) \ \ul{\psi}_\lam \ \ .
\eeqa

Analogously to Lemma \ref{lemext0} in the euclidean case, we 
obtain for 
each state $\xi$ on $\9F_\8b(\7R_+)$ a strongly continuous representation 
$\ul{U}_\xi$ of the Poincar\'e group as well as 
a representation $\ul{\pi}_\xi$ 
of the scaling algebra $\ul{\6A}$ on $\ul{\2H}_\xi$ such that 
\beqa
\ul{\pi}_\xi(\al_g\1a)&=&\ul{U}_\xi(g)\ul{\pi}_\xi(\1a)\ul{U}_\xi(g)^* 
\eeqa
where $\ul{\pi}_\xi$ and $\ul{U}_\xi$ are given by
\beqa
\ul{\pi}_\xi(\1a)\1q_\xi[\ul{\psi}]
&=&\int^\oplus \8d\xi(\lam) \ \1a(\lam)\ul{\psi}_\lam
\vs\vs
\ul{U}_\xi(g)\1q_\xi[\ul{\psi}]
&=&\int^\oplus \8d\xi(\lam) \ U_\lam(g) \ \ul{\psi}_\lam  \ \ .
\eeqa

We are now prepared to formulate a lemma which turns out to be
very useful for our subsequent analysis and which relates 
the representation $\ul{\pi}_\zeta$ to the GNS triple 
$(\2H_{\ul{\om}_\zeta},\pi_{\ul{\om}_\zeta},\Om_{{\ul{\om}_\zeta}})$
of the scaling limit $\ul{\om}_\zeta$:

\blem\label{leminter}
For each limit functional $\zeta$ there exists an isometry 
\beqa
\1v_\zeta:\2H_{\ul{\om}_\zeta}\to \ul{\2H}_\zeta
\eeqa
which intertwines the  
GNS representation $\pi_{\ul{\om}_\zeta}$ of 
the scaling limit $\ul{\om}_\zeta$ and the representation 
of $\ul{\pi}_\zeta$ on the scaling algebra:
\beqa
\1v_\zeta\pi_{\ul{\om}_\zeta}(\1a)&=&\ul{\pi}_\zeta(\1a)\1v_\zeta
\eeqa
for each $\1a\in\ul{\6A}$.
\elem
\bpr
An isometry 
\beqa
\1v_\zeta:\2H_{\ul{\om}_\zeta}\to \ul{\2H}_\zeta
\eeqa
is given by the prescription 
\beqa
\1v_\zeta[\pi_{\ul{\om}_\zeta}(\1a)\Om_{{\ul{\om}_\zeta}}]
&=&\ul{\pi}_\zeta(\1a)\ul{\Om}_\zeta
\eeqa
where $\ul{\Om}_\zeta$ is the equivalence class in $\ul{\2H}_\zeta$ of 
the function $\ul{\Om}:\lam\mapsto\Om=\Psi[\11]$.
Indeed we have 
\beqa
\|\1v_\zeta[\pi_{\ul{\om}_\zeta}(\1a)\Om_{{\ul{\om}_\zeta}}]\|^2
&=&
\<\ul{\pi}_\zeta(\1a)\ul{\Om}_\zeta,\ul{\pi}_\zeta(\1a)\ul{\Om}_\zeta\>
\vs\vs
&=&
\int \8d\zeta(\lam) \ \<\Om,\1a(\lam)^*\1a(\lam)\Om\>
\vs\vs
&=&
\<\ul{\om}_\zeta,\1a(\lam)^*\1a(\lam)\>
\eeqa
and $\1v_\zeta$ is a well defined isometry which intertwines 
the representations 
$\ul{\pi}_\zeta$ and $\pi_{\ul{\om}_\zeta}$.
\epr

\section{The proof of Theorem \ref{the1}}
\label{app2}
\paragraph{\em Construction of an intertwining isometry.}
We have shown in the previous paragraph that the 
Hilbert space $\ul{\2H}_\zeta$ carries a faithful 
representation of the scaling algebra 
$\ul{\6A}_\zeta$. 
On the other hand, the Hilbert space $\2H_\zeta$ carries a faithful 
representation of the algebra  
$\6A_\zeta$ which is constructed from the scaling limit 
$(\ul{\6B},\ul{\be},\ul{\eta}_\zeta)$ of the euclidean field 
$(\6B,\be,\eta)$.
Within this paragraph we construct an isometry from 
$\2H_\zeta$ to $\ul{\2H}_\zeta$. It turns out that 
this particular isometry induces an algebra isomorphism 
between $\6A_\zeta$ and $\ul{\6A}_\zeta$.

For each $s\in\7R_+$ and for each $\1b\in\ul{\6B}(s,e)$
and for each time zero operator $\1b_0\in\hat{\ul{\6B}}(e^\perp)$  
we define bounded operators 
$\ul\pi_{(s,\zeta)}(\1b)$ and $\ul{\pi}_\zeta(\1b_0)$ on 
$\ul{\2H}_\zeta$ according to 
\beqa
\ul\pi_{(s,\zeta)}(\1b)\1q_\zeta[\ul{\psi}]
&=&\int^\oplus \8d\zeta(\lam) \ 
\pi_s(\1b(\lam))\ul{\psi}_\lam
\vs\vs
\ul\pi_\zeta(\1b_0)\1q_\zeta[\ul{\psi}]
&=&\int^\oplus \8d\zeta(\lam) \ \pi(\1b_0(\lam))\ul{\psi}_\lam
\eeqa
for each $\ul{\psi}\in\ul{\2H}$.

\blem\label{lemiso}
There exists an isometry 
\beqa
\1u_\zeta:\2H_\zeta&\to &\ul{\2H}_\zeta
\eeqa
such that for each $s\in\7R_+$ and for each $\1b\in\ul{\6B}(s,e)$
and for each time zero operator $\1b_0\in\hat{\ul{\6B}}(e^\perp)$  
the identities 
\beqa
\1u_\zeta\pi_{(s,\zeta)}(\1b)&=&\ul{\pi}_{(s,\zeta)}(\1b)\1u_\zeta
\vs\vs
\1u_\zeta\pi_\zeta(\1b_0)&=&\ul{\pi}_\zeta(\1b_0)\1u_\zeta
\eeqa
are valid.
\elem
\bpr
We define the operator $\1u_\zeta$ by according to 
\beqa
\1u_\zeta\Psi_\zeta[\1b]&:=&\ul{\Psi}_\zeta[\1b]
\eeqa
where $\ul{\Psi}[\1b]\in\ul{\2H}$ is defined by 
\beqa
\ul{\Psi}[\1b](\lam)&:=&\Psi[\1b(\lam)] 
\eeqa
and $\ul{\Psi}_\zeta[\1b]$ is the corresponding equivalence class in 
$\ul{\2H}_\zeta$.
Indeed, $\1u_\zeta$ is a well defined isometry, since 
we have 
\beqa
\|\1u_\zeta\Psi_\zeta[\1b]\|^2&:=&
\<\ul{\Psi}[\1b],\ul{\Psi}[\1b]\>_\zeta
\vs\vs
&=&\int\8d\zeta(\lam)\ \<\eta,j_e(\1b(\lam))\1b(\lam)\>
\vs\vs
&=&\<\ul{\eta}_\zeta,j_e(\1b)\1b\>
\vs\vs
&=&\|\Psi_\zeta[\1b]\|^2 \ \ .
\eeqa
We compute for each $s\in\7R_+$, for each $\1b\in\ul{\6B}(s,e)$
and for each $\1b_1\in\ul{\6B}(e)$:
\beqa
\1u_\zeta\pi_{(s,\zeta)}(\1b)\Psi_\zeta[\1b_1]
&=&
\1u_\zeta\Psi_\zeta[\1b\ul{\be}_{se}\1b_1]
\vs\vs
&=&
\ul{\Psi}_\zeta[\1b\ul{\be}_{se}\1b_1]
\vs\vs
&=&
\ul{\pi}_{(s,\zeta)}(\1b) \ul{\Psi}_\zeta[\1b_1]
\vs\vs
&=&
\ul{\pi}_{(s,\zeta)}(\1b)\1u_\zeta\Psi_\zeta[\1b_1] \ \ .
\eeqa
The identity 
\beqa
\1u_\zeta\pi_\zeta(\1b_0)&=&\ul{\pi}_\zeta(\1b_0)\1u_\zeta
\eeqa
follows from a similar computation which implies the lemma. 
\epr

\blem\label{lemhelp1}
The isometry $\1u_\zeta$ intertwines the representations 
$U_\zeta$ and $\ul{U}_\zeta$: For each Poincar\'e
transformation $g$ the identity  
\beqa
\1u_\zeta U_\zeta(g)
&=&\ul{U}_\zeta(g)\1u_\zeta
\eeqa
holds true.
\elem
\bpr
Let $e_1$ be a euclidean direction, perpendicular to $e$.
and let $\be_{(e,e_1)}$ 
be the 
one-parameter group of automorphisms which are given by the 
rotations in the $e,e_1$ plane.
Let $\Gam(e,r)$, $r>0$, be the open cone which is invariant under the 
stabilizer subgroup of $e$ and which has opening angle 
$\pi/2-r$ with $r\in(0,\pi/2)$.

According to \cite{FrohOstSeil}, for each $\tau$, with $|\tau|\leq r$,
there exist self adjoint operators
\beqa 
V_{(e,e_1)}(\tau):\Psi[\6B(\Gam(e,r))]&\to& \2H
\vs\vs
V_{(e,e_1,\zeta)}(\tau):\Psi_\zeta[\ul{\6B}(\Gam(e,r))]&\to& \2H_\zeta
\eeqa
which are given by 
\beqa
V_{(e,e_1)}(\tau)\Psi[b]&:=&\Psi[\be_{(e,e_1,\tau)}b]
\vs\vs
V_{(e,e_1,\zeta)}(\tau)\Psi_\zeta[\1b]
&:=&\Psi_\zeta[\ul{\be}_{(e,e_1,\tau)}\1b]
\ \ .
\eeqa
We identify the hyperplane $e^\perp$ with a space like 
hyperplane in Minkowski space and $e$ with the corresponding timelike 
direction. Let $B_{(e,e_1)}$ and $B_{(e,e_1,\zeta)}$ 
be the (anti-selfadjoint) generators of the Lorentz boosts
in $e,e_1$ direction within the representation $U$ and $U_\zeta$
respectively. Then the identities 
\beqa
V_{(e,e_1)}(\tau)&=&\exp(\8i\tau B_{(e,e_1)})
\vs\vs
V_{(e,e_1,\zeta)}(\tau)&=&\exp(\8i\tau B_{(e,e_1,\zeta)})
\eeqa
are valid. For operators 
$\1b_0\in\ul{\6B}(e)$ and $\1b_1\in\ul{\6B}(\Gam(e,r))$, 
we introduce complex functions by 
\beqa
\ul{F}^{[e,e_1|\1b_0,\1b_1]}_\lam(z)&:=&
\<\Psi[\1b_0(\lam)],\exp(z B_{(e,e_1)})\Psi[\1b_1(\lam)]\>
\vs\vs
F^{[e,e_1|\1b_0,\1b_1]}_\zeta(z)&:=&
\<\Psi_\zeta[\1b_0],\exp(z B_{(e,e_1,\zeta)})\Psi_\zeta[\1b_1]\> \ \ .
\eeqa
According to the analysis, carried out in \cite{FrohOstSeil,Schl97},
the functions $\ul{F}^{[e,e_1|\1b_0,\1b_1]}_\lam$, $\lam>0$, and 
$F^{[e,e_1|\1b_0,\1b_1]}_\zeta$ 
are holomorphic in the open strip $\7R+\8i(-r,r)$.

Furthermore, let $H_e$ and $H_{e,\zeta}$ be the positive generators 
(Hamilton operators) of the 
semi group of contractions $V_e$ and $V_{(e,\zeta)}$ respectively, given by 
\beqa
V_e(\tau)\Psi[b]&:=&\Psi[\be_{\tau e}b]
\vs\vs
V_{(e,\zeta)}(\tau)\Psi_\zeta[\1b]&:=&\Psi_\zeta[\ul{\be}_{\tau e}\1b]
\eeqa
for each $\tau>0$ and for each $b\in\6B(e)$ and for each 
$\1b\in\ul{\6B}(e)$.
For $\1b_0,\1b_1\in\ul{\6B}(e)$ 
we introduce again complex functions
\beqa
\ul{F}^{[e|\1b_0,\1b_1]}_\lam(z)&:=&
\<\Psi[\1b_0(\lam)],\exp(\lam z H_e)\Psi[\1b_1(\lam)]\>
\vs\vs
F^{[e|\1b_0,\1b_1]}_\zeta(z)&:=&
\<\Psi_\zeta[\1b_0],\exp(z H_{(e,\zeta)})\Psi_\zeta[\1b_1]\> \ \ .
\eeqa
which are holomorphic in the upper half plane $\7R+\8i\7R_+$.

The lemma follows mainly from the subsequent statements, which are proven 
in the next section Appendix \ref{app3}:

\bslem\label{slem1}
For each limit functional 
$\zeta\in\6S[\8{C}(\7R_+)]$,
for each $\1b_0\in\ul{\6B}(e)$, and for each $\1b_1\in\ul{\6B}(\Gam(e,r))$:
\bdes 

\itno 1
The prescription
\beqa
\ul{F}^{[e,e_1|\1b_0,\1b_1]}_\zeta:
z&\mapsto&\int\8d\zeta(\lam) \ \ul{F}^{[e,e_1|\1b_0,\1b_1]}_\lam(z)
\eeqa
yields a well defined function which is holomorphic in the strip
$\7R+\8i(-r,r)$.

\itno 2
The prescription
\beqa
\ul{F}^{[e|\1b_0,\1b_1]}_\zeta:
z&\mapsto&\int \8d\zeta(\lam) \ \ul{F}^{[e|\1b_0,\1b_1]}_\lam(z)
\eeqa
yields a well defined function which is holomorphic in the upper
half plane $\7R+\8i\7R_+$.

\edes

\eslem

\bslem\label{slem2}
For each limit functional 
$\zeta\in\6S[\8{C}(\7R_+)]$,
for each $\1b_0\in\ul{\6B}(e)$, and for each $\1b_1\in\ul{\6B}(\Gam(e,r))$
the identities
\beqa
\ul{F}^{[e,e_1|\1b_0,\1b_1]}_\zeta&=&F^{[e,e_1|\1b_0,\1b_1]}_\zeta
\vs\vs
\ul{F}^{[e|\1b_0,\1b_1]}_\zeta&=&F^{[e|\1b_0,\1b_1]}_\zeta
\eeqa
holds true.
\eslem

Let $e_1\perp e$ and let $g$ be the Lorentz transformation such that 
\beqa
U(g)&=&\exp(tB_{(e,e_1)}) \ \ .
\eeqa
According to Sublemma \ref{slem2} we compute 
\beqa
\<\Psi_\zeta[\1b_0],\1u_\zeta^*\ul{U}_\zeta(g)
\ul{\Psi}_\zeta[\1b_1]\>
&=&
\ul{F}^{[e,e_1|\1b_0,\1b_1]}_\zeta(t)
\vs\vs
&=&
F^{[e,e_1|\1b_0,\1b_1]}_\zeta(t)
\vs\vs
&=&
\<\Psi_\zeta[\1b_0],U_\zeta(g)\Psi_\zeta[\1b_1]\>
\eeqa
and since $\Psi_\zeta[\ul{\6B}(e)]$ is dense in $\2H_\zeta$,
we conclude that 
\beqa
\ul{U}_\zeta(g)\ul{\Psi}_\zeta[\1b_1]
&=&\1u_\zeta U_\zeta(g)\Psi_\zeta[\1b_1]
\eeqa
which implies
\beqa
\1u_\zeta U_\zeta(g)&=&\ul{U}_\zeta(g) \1u_\zeta
\eeqa
for each Lorentz boost $g$. Analogously we conclude for the 
time like translations in $e$ direction  
\beqa
\<\Psi_\zeta[\1b_0],\1u_\zeta^*\ul{U}_\zeta(te)
\ul{\Psi}_\zeta[\1b_1]\>
&=&
\ul{F}^{[e|\1b_0,\1b_1]}_\zeta(t)
\vs\vs
&=&
F^{[e|\1b_0,\1b_1]}_\zeta(t)
\vs\vs
&=&
\<\Psi_\zeta[\1b_0],U_\zeta(te)\Psi_\zeta[\1b_1]\>
\eeqa
and thus 
\beqa
\1u_\zeta U_\zeta(te)&=&\ul{U}_\zeta(te) \1u_\zeta
\eeqa
for each $t\in\7R$.

Now, let $g$ be an element of the stabilizer subgroup 
of the hyperplane $e^\perp$, then we
compute for each $\1b\in\ul{\6B}(e)$:
\beqa
\1u_\zeta U_\zeta(g)\Psi_\zeta[\1b]
&=&
\1u_\zeta\Psi_\zeta[\ul{\be}_{g}\1b]
\vs\vs
&=&
\ul{\Psi}_\zeta[\ul{\be}_{g}\1b]
\vs\vs
&=&
\ul{U}_\zeta(g)\ul{\Psi}_\zeta[\1b] \ \ .
\eeqa
Therefore the identity 
\beqa
\1u_\zeta U_\zeta(g)&=&\ul{U}_\zeta(g) \1u_\zeta
\eeqa
is valid for each Poncar\'e transformation $g$ which completes the 
proof.
\epr

\paragraph{\em Proof of Theorem \ref{the1}.}
By Lemma \ref{leminter} 
the quantum field $(\ul{\6A}_\zeta,\ul{\al}_\zeta,\ul{\om}_\zeta)$
may be identified with
the scaling algebra $(\ul{\6A},\ul{\al})$ in the 
vacuum representation $\ul{\pi}_\zeta$.

Recall, that the scaling algebra $\ul{\6A}$ can be introduced in   
terms of the time-zero algebras 
$\hat{\ul{\6B}}(\9V)$, where $\9V$ is a bounded and convex region 
in the time slice $e^\perp$.
For a double cone $\9O$, the local
scaling algebra $\ul{\6A}(\2O)$ is generated by all functions 
\beqa
\ul\Pi[f,\1b]:\lam&\mapsto&
\int \8dg \ f(g) \ U_\lam(g)\pi(\1b(\lam))U_\lam(g)^*
\eeqa
where $f\in\9C^\infty_0(\Poin)$ is a smooth function on the 
Poincar\'e group with compact support 
such that  $g\9V\subset\9O$ for each $g$ in the support of $f$ and   
$\1b\in\hat{\ul{\6B}}(\9V)$.
Applying the representation $\ul{\pi}_\zeta$ to 
$\ul\Pi[f,\1b]$ yields
\beqa
\ul{\Pi}_\zeta[f,\1b]&:=&\ul{\pi}_\zeta(\ul\Pi[f,\1b])
\ \  = \ \ 
\int \8dg \ f(g) \ \ul{U}_\zeta(g)\ul{\pi}_\zeta(\1b)\ul{U}_\zeta(g)^* \ \ .
\eeqa

On the other hand, 
for a double cone $\9O$, the local algebra $\6A_\zeta(\9O)$ 
of the quantum field $(\6A_\zeta,\al_\zeta,\om_\zeta)$ 
which can be constructed from the 
euclidean field $(\ul{\6B},\ul{\be},\ul{\eta}_\zeta)$, is generated  
by operators of the form 
\beqa
\Pi_\zeta[f,\1b]&=& \int \8dg \ f(g) \ U_\zeta(g)\pi_\zeta(\1b)U_\zeta(g)^*
\eeqa
with a time-zero operator $\1b\in\hat{\ul{\6B}}(\9V)$
and a smooth function $f\in\9C^\infty_0(\Poin)$, 
such that  $g\9V\subset\9O$ for each $g$ in the support of $f$.

According to the intertwining properties of the isometry $\1u_\zeta$ 
we conclude form Lemma \ref{lemiso} and from Lemma \ref{lemhelp1}
that the identity 
\beqa
&&\1u_\zeta\biggl[\int\8dg \ f(g) \ U_\zeta(g)
\pi_{(s,\zeta)}(\1b)U_\zeta(g)^*\biggr]\1u_\zeta^*
\vs\vs
&=&
\int\8dg \ f(g) \ \1u_\zeta U_\zeta(g)
\pi_{(s,\zeta)}(\1b)U_\zeta(g)^*\1u_\zeta^*
\vs\vs
&=&
\int\8dg \ f(g) \ \ul{U}_\zeta(g)
\ul{\pi}_{(s,\zeta)}(\1b)\ul{U}_\zeta(g)^*
\eeqa
is valid for each $s\in\7R_+$.
This implies 
\beqa
\1u_\zeta\Pi_\zeta[f,\1b]\1u_\zeta^* 
&=&\ul{\Pi}_\zeta[f,\1b]
\eeqa
for each  smooth function $f\in\9C^\infty_0(\Poin)$ and for each 
time-zero operator $\1b\in\hat{\ul{\6B}}(e^\perp)$. 
Therefore, we get 
\beqa
\1u_\zeta\6A_\zeta(\2O)\1u_\zeta^*&:=&\ul{\pi}_\zeta(\ul{\6A}(\2O))
\ \ = \ \ \ul{\6A}_\zeta(\2O)
\eeqa
and the map  
\beqa
\iota_\zeta:\6A_\zeta\to\ul{\6A}_\zeta &;& \1a\mapsto \1u_\zeta\1a\1u_\zeta^*
\eeqa
yields an isomorphism of the quantum fields
$(\6A_\zeta,\al_\zeta,\om_\zeta)$ and 
$(\ul{\6A}_\zeta,\ul{\al}_\zeta,\ul{\om}_\zeta)$ since 
\beqa
\iota_\zeta\circ\al_\zeta&=&\ul{\al}_\zeta\circ\iota_\zeta
\eeqa
is valid according to Lemma \ref{lemhelp1} and 
\beqa
\om_\zeta&=&\ul{\om}_\zeta\circ \iota_\zeta
\eeqa
holds true, which is a consequence of the fact that 
$\1u_\zeta\Om_\zeta=\ul{\Om}_\zeta$, where 
$\ul{\Om}_\zeta$ is the equivalence class in $\ul{\2H}_\zeta$  
of the constant function $\lam\mapsto\Om$.
\epr
  
\section{Remarks on holomorphic functions}
\label{app3}
The statements of 
Sublemma \ref{slem1} and Sublemma \ref{slem2} can directly be 
obtained form some general statement on 
holomorphic functions, 
which we discuss in the subsequent.

\blem\label{montel}
Let $I\subset\7R$ be an open connected subset and let 
$f_\lam\in\2O(\7R+\8iI)$, $\lam\in\7R_+$, be a family of 
functions which are holomorphic in $\7R+\8iI$.
If there exists a constant $K>0$ such that 
the bound
\beqa
|f_\lam(z)| &\leq&K
\eeqa
holds true for each $\lam\in\7R_+$ and for each $z\in\7R+\8iI$,
then for each limit functional $\zeta\in\6S[\8C(\7R_+)]$
there exists a function $f_\zeta\in\2O(\7R+\8iI)$, holomorphic in $\7R+\8iI$,
which is uniquely determined by the prescription
\beqa
f_\zeta(z)&=&\int\8d\zeta(\lam) \ f_\lam(z)  \ \ .
\eeqa 
\elem
\bpr
The lemma is nothing else but Montel's Theorem expressed in 
terms of limit functionals. 
According to our assumption the family of holomorphic functions 
$f_\lam\in\2O(\7R+\8iI)$, 
$\lam\in\7R_+$, is uniformly bounded by a constant $K$, i.e.
\beqa
|f_\lam(z)| &\leq&K
\eeqa
for each $\lam\in\7R_+$ and for each $z\in\7R+\8iI$.
Let $r>0$ and let $D_r(z)$ be the closed disc 
in $\7C$ with radius $r$ and center $z$.
For $z\in\7R+\8iI$ we choose $r>0$ with $D_r(z)\subset\7R+\8iI$.
For $z_1,z_2\in D_r(z)$ we immediately get the estimate
\beqa
|f_\lam(z_1)-f_\lam(z_2)|&=&\biggm|
\int_{z_1}^{z_2}\8dz f_\lam'(z)\biggm|  
\ \ \leq  \ \ |z_1-z_2|4r^{-1} K
\eeqa
uniformly in $\lam$. This implies, since $\zeta$ is a positive functional, 
\beqa
|f_\zeta(z_1) - f_\zeta(z_2)| &\leq& |z_1-z_2| 4 r^{-1} K
\eeqa
for each $z_1,z_2\in D_r(z)$. Now, the function 
\beqa
f_\zeta:z&\mapsto& \int\8d\zeta(\lam) \ f_\lam(z)
\eeqa
is integrable with respect to the natural measure on the circle 
$\pa D_r(z)$ since it is continuous in $\7R+\8iI$ 
(in particular uniformly continuous in $D_r(z)$ for each $z\in\7R+\8iI$).
On the other hand, the function 
\beqa
f_{\pa D_r(z)}:\lam\mapsto \int_{\pa D_r(z)} \8dz' \ f_\lam(z')
\eeqa
is uniformly bounded in $\lam$, i.e. $f_{\pa D_r(z)}$ is 
contained in $\9F_\8b(\7R_+)$, and hence $f_{\pa D_r(z)}$ is  
measurable with respect to $\zeta$.
Thus Fubini's theorem can be applied which states that the integration 
over $\pa D_r(z)$ and the integration with respect to the measure
$\zeta$ can be exchanged. This gives 
\beqa
\int_{\pa D_r(z)} \8dz' \ f_\zeta(z')
&=& \int_{\pa D_r(z)} \8dz' \int\8d\zeta(\lam) \ f_\lam(z')
\vs\vs
&=&
\int\8d\zeta(\lam) \ \int_{\pa D_r(z)} \8dz' \ f_\lam(z')
\vs\vs
&=&0 
\eeqa 
which implies that $f_\zeta$ is holomorphic in $z$.
\epr

\paragraph{\it Proof of Sublemma \ref{slem1}.}
In order to prove Sublemma \ref{slem1} we show that the 
functions
\beqa
\ul{F}^{[e,e_1|\1b_0,\1b_1]}_\lam(z)&=&
\<\Psi[\1b_0(\lam)],\exp(z B_{(e,e_1)})\Psi[\1b_1(\lam)]\>
\vs\vs
F^{[e|\1b_0,\1b_1]}_\lam(z)&=&
\<\Psi_\zeta[\1b_0],\exp(\lam z H_e)\Psi_\zeta[\1b_1]\> 
\eeqa
are uniformly bounded. Since the operators 
$\exp(z B_{(e,e_1)})$ and $\exp(\lam z H_e)$ are unitary 
for real $z$ and since the norm of the vector $\Psi[b]\in\2H$, 
$b\in\6B(e)$, is bounded by the operator norm $\|b\|$ we conclude 
the estimate
\beqa
|\ul{F}^{[e,e_1|\1b_0,\1b_1]}_\lam(z)|&\leq&\|\1b_0\| \ \|\1b_1\|
\vs\vs
|\ul{F}^{[e|\1b_0,\1b_1]}_\lam(z)|&\leq&\|\1b_0\| \ \|\1b_1\|
\eeqa
for each $z$ in the corresponding region of holomorphy and 
for each $\lam\in\7R_+$.
Thus Lemma \ref{montel} can be applied, which proves
Sublemma \ref{slem1}.
\epr

A further well known fact concerning holomorphic function is the following:
\blem\label{ident}
$f,\ul{f}\in\2O(\7R+\8iI)$, be two
functions which are holomorphic in $\7R+\8iI$, where $I$ is connected.
If $f$ and $\ul{f}$ coincide within the imaginary points $\8iI$, 
then $f=\ul{f}$.
\elem

\paragraph{\it Proof of Sublemma \ref{slem2}.}
For each $\1b_1\in\ul{\6B}(\Gam(e,r))$, 
the functions $\ul{F}^{[e,e_1|\1b_0,\1b_1]}_\zeta$ and
$F^{[e,e_1|\1b_0,\1b_1]}_\zeta$ are holomorphic in the 
open strip $\7R+\8i(-r,r)$ and one easily computes 
for each $\tau\in (-r,r)$ 
\beqa
\ul{F}^{[e,e_1|\1b_0,\1b_1]}_\zeta(\8i\tau)&=&
\int \8d\zeta(\lam) \
\<\Psi[\1b_0(\lam)],\exp(\8i\tau B_{(e,e_1)})\Psi[\1b_1(\lam)]\>
\vs\vs
&=&
\int \8d\zeta(\lam) \
\<\Psi[\1b_0(\lam)],\Psi[\ul{\be}_{(e,e_1,\tau)}\1b_1(\lam)]\>
\vs\vs
&=&
\int \8d\zeta(\lam) \
\<\eta,j_e(\1b_0(\lam))\ul{\be}_{(e,e_1,\tau)}\1b_1(\lam)]\>
\vs\vs
&=&
\<\eta_\zeta,j_e(\1b_0)\ul{\be}_{(e,e_1,\tau)}\1b_1\>
\vs\vs
&=&
\<\Psi_\zeta[\1b_0],\exp(\8i\tau B_{(e,e_1,\zeta)})\Psi_\zeta[\1b_1]\>]\>
\vs\vs
&=&
F^{[e,e_1|\1b_0,\1b_1]}_\zeta(\8i\tau)
\eeqa
which implies that $F^{[e,e_1|\1b_0,\1b_1]}_\zeta$ and
$\ul{F}^{[e,e_1|\1b_0,\1b_1]}_\zeta$ coincide in the imaginary points and 
by Lemma \ref{ident} it follows that 
$F^{[e,e_1|\1b_0,\1b_1]}_\zeta=\ul{F}^{[e,e_1|\1b_0,\1b_1]}_\zeta$.
Analogously one proves that
$F^{[e|\1b_0,\1b_1]}_\zeta=\ul{F}^{[e|\1b_0,\1b_1]}_\zeta$ 
holds true also.
\epr
\end{appendix}
\newpage


\end{document}